\documentclass{aa}  
\usepackage{natbib}
\usepackage[FIGTOPCAP]{subfigure}

\usepackage{graphicx}
\usepackage{color}
\usepackage[table, dvipsnames]{xcolor}
\usepackage{hyperref}
\hypersetup{
    colorlinks   = true,
    citecolor=blue,
    filecolor=black,
    linkcolor=blue,
    urlcolor=blue
}

\usepackage[flushleft]{threeparttable}
\usepackage{multirow}
\usepackage{tablefootnote}
\usepackage{scrextend}

\usepackage{txfonts}
\def\hii{H\,{\sc ii}}
\def\Msun{M$_\odot$}
\def\Lsun{L$_\odot$}
\def\xsh{X-shooter}
\def\prodimo{{\sc ProDiMo}}

\defcitealias{2017A&A...604A..78R}{RT17}

\begin{document}

   \title{Massive pre-main-sequence stars in M17}

   \subtitle{Modelling hydrogen and dust in MYSO disks}

   \author{\mbox{F. Backs}\inst{1}
          \and
          \mbox{J. Poorta}\inst{1}
          \and
          \mbox{Ch. Rab}\inst{4,5}
          \and
          \mbox{A.R. Derkink}\inst{1}
          \and
          \mbox{A. de Koter}\inst{1,2}
          \and
          \mbox{L. Kaper}\inst{1}
          \and
          \mbox{M.C. Ram\'irez-Tannus}\inst{3}
          \and
          \mbox{I. Kamp}\inst{6}
          }

   \institute{Anton Pannekoek Institute for Astronomy, University of Amsterdam, Science Park 904, 1098 XH Amsterdam, The Netherlands, email: \texttt{f.p.a.backs@uva.nl}
   \and
   Institute of Astrophysics, Universiteit Leuven, Celestijnenlaan 200 D, 3001 Leuven, Belgium 
   \and
   Max Planck Institute for Astronomy, Königstuhl 17, D-69117 Heidelberg, Germany   
   \and
   Universit\"{a}ts-Sternwarte, Fakult\"{a}t f\"{u}r Physik, Ludwig-Maximilians-Universit\"{a}t M\"{u}nchen, Scheinerstr. 1, 
   81679 M\"{u}nchen, Germany   
   \and
   Max-Planck-Institut f\"{u}r extraterrestrische Physik, Giessenbachstrasse 1, 85748 Garching, Germany
   \and
   Kapteyn Astronomical Institute, University of Groningen, Postbus 800, 9700 AV Groningen, The Netherlands
             }

   \date{Received 26 August, 2022; accepted 20 December, 2022}

 
  \abstract
   {The young massive-star-forming region M17 contains optically visible massive pre-main-sequence stars that are surrounded by circumstellar disks. Such disks are expected to disappear when these stars enter the main sequence. The physical and dynamical structure of these remnant disks are poorly constrained, especially the inner regions where accretion, photo-evaporation, and companion formation and migration may be ongoing.} %
   {We aim to constrain the physical properties of the inner parts of the circumstellar disks of massive young stellar objects B243 (6\,M$_\odot$) and B331 (12\,M$_\odot$), two systems for which the central star has been detected and characterized previously despite strong dust extinction. }
   {Two-dimensional radiation thermo-chemical modelling with \prodimo\ of double-peaked hydrogen lines of the Paschen and Brackett series observed with \xsh\ was used to probe the properties of the inner disk of the target sources. The model was modified to treat these lines. Additionally, the dust structure was studied by fitting the optical and near-infrared spectral energy distribution. 
   }
   {B243 features a hot gaseous inner disk with dust at the sublimation radius at $\sim$3\,AU. The disk appears truncated at roughly 6.5\,AU; a cool outer disk of gas and dust may be present, but it cannot be detected with our data. B331 also has a hot gaseous inner disk. A gap separates the inner disk from a colder dusty outer disk starting at up to $\sim$100\,AU. In both sources the inner disk extends to almost the stellar surface. Chemistry is essential for the ionization of hydrogen in these disks.}
   {The lack of a gap between the central objects and these disks suggests that they accrete through boundary-layer accretion. This would exclude the stars having a strong magnetic field. 
   Their structures suggest that both disks are transitional in nature, that is to say they are in the process of being cleared, either through boundary-layer accretion, photo-evaporation, or through companion activity.}


   \maketitle
%

\section{Introduction}

The evolution of high-mass protostars and their descendants, the massive young stellar bbjects (MYSOs), is expected to be very fast, $\sim$10$^5$\,yr \citep[e.g.][]{2009ApJ...691..823H}, and, until very late in the build-up process, hidden from view as it unfolds deep within dusty natal clouds. 
Though much is still unclear about the mechanism that leads to the assembly of a massive star, most theories agree on the need for a dense and massive accretion disk 
\citep[e.g.][]{2014prpl.conf..149T,2016A&ARv..24....6B}, possibly in combination with bipolar polar outflows at relatively early phases of formation.  
Likely, the massive disks that develop are complex in structure with multiple processes acting to cause instabilities, for example magneto-rotational instabilities leading to turbulence and associated angular momentum redistribution \citep[e.g.][]{1998RvMP...70....1B} 
and gravitational instabilities giving rise to the formation of companions \citep[e.g.][]{2010ApJ...708.1585K,2020A&A...644A..41O}. 

The inner parts of the disk are particularly interesting for at least two reasons. First, from these regions the accretion of mass and redistribution of angular momentum takes place. Important open problems are the gas content of the inner disk reservoir and the accretion efficiency of this material close to the arrival of the star on the main sequence. The assembly of angular momentum is particularly interesting in view of the buildup and outcome of the surface rotation velocity, the latter property being key to reaching a new level of understanding of main sequence single and binary star evolution \citep{2022NatAs...6..480W}. Second, the mechanism(s) disrupting the circumstellar disk likely act first fairly close to the central star. The main candidates for this process are disk winds, outflows resulting for example from photo-evaporation \citep[e.g.][]{2009ApJ...690.1539G,2010MNRAS.401.1415O,2020A&A...643A..32G}, stellar winds, and stellar or planetary companion formation \citep[e.g.][]{2013A&A...560A..40M}. Our understanding of disks that are in the process of clearing themselves out (so-called transitional disks) stems mostly from low- and intermediate-mass stars, where the process takes 10-20\% of the total YSO lifetime \citep{2011ApJS..195....3F}. Given the short duration of the MYSO phase \citep[$10^{4}-10^{5}$\,yr;][]{2009ApJ...691..823H}, it remains to be seen whether such a modest fraction of the MYSO lifetime will suffice, or whether perhaps the majority or even all MYSO disks are transitional in nature.

Atomic recombination and CO bandhead emission lines are the main (gas-phase) diagnostics of the properties, structure, and kinematics of the inner dust-free regions of the disks of MYSOs. The hydrogen line-formation mechanism near the end of the MYSO phase is complex, as the stars are likely not accreting enough to cause significant hydrogen ionization as a result of viscous heating and shocks, and the central stars' photospheres are not hot enough to produce sufficient ionizing radiation and associated hydrogen photo-ionization.

Evidence is mounting that the hydrogen emission originates from very close to the stellar surface. \citet{2021A&A...654A.109K} present an interferometric K-band study of a sample of MYSOs spatially resolving the dust and Br$\gamma$ emission in objects ranging from 9 to 16~M$_\odot$. They find Br$\gamma$ emission to originate from $\sim$0.9 to 6~AU from the central star and the 2.2~$\mu$m emission to originate from $\sim$1.4 to 6.8~AU from the central star depending on the object. They conclude that the hydrogen emission originates from a more compact region than the 2.2~$\mu$m dust emission, with the continuum dust emission being consistent with originating from the dust sublimation radius. This is in line with the findings of \citet{2016A&A...589L...4C} who find a Br$\gamma$ emitting region of 6--13 AU and a further out continuum emitting region of 17 AU for a 20~M$_\odot$ MYSO. Additionally, they detected an outflow in Br$\gamma$. \citet{2010Natur.466..339K} also find a dust free cavity analogous to those found in lower mass YSOs compatible with the dust sublimation radius. \citet{2020A&A...635L..12G} detect CO emission in the inner gaseous disk of the massive (15\,$M_{\odot}$) YSO NGC\,2021\,IRS\,2, which is well within the dust sublimation radius. The dynamical nature of this hot gas close to the central star is still a matter of debate. MYSOs could have weak or absent magnetic fields similar to their main-sequence counterparts in which few stars are observed to have strong magnetic fields \citep[e.g.][]{2015A&A...582A..45F,2016MNRAS.456....2W}. Thus accretion might proceed through boundary layer (BL) accretion rather than magnetospheric accretion. Therefore, disks might potentially extend all the way to the stellar surface, becoming dust free inside of the dust sublimation radius. Alternatively, the nearby gas could result from bipolar outflows \citep{2021A&A...648A..62F}. 

On the basis of infrared spectroscopy, photometry, and high-spatial resolution imaging of optical scattered light from small dust grains, disks around YSOs have been classified into two groups \citep[e.g.][]{2017A&A...603A..21G}, reflecting large-scale morphological properties. Group\,{\sc I} sources show large extended
disks with a pronounced gap in the centre. This results in low or lacking near-infrared excess due to the absence of hot dust. Such inner dust-free cavities extending beyond the dust sublimation radius are observed in MYSOs \citep{2021A&A...648A..62F}.  Group\,{\sc II} sources show strong NIR-excess, but limited excess at longer wavelengths. This suggests the presence of hot dust near the star, but the absence of cooler dust further out, which can be caused by a small truncated disk, or a self-shadowed disk. Though this classification does clearly imply a different structure of the inner disk (where the hot dust is located), it does not link to a potential gas reservoir in these inner regions. Scrutinizing this hot gas, if present, may help to understand the processes shaping the inner disk regions. 

In this paper we present the modelling of hydrogen emission lines and dust continuum emission from two MYSOs using the radiation thermo-chemical code \prodimo\ \citep{Woitke09, 2010A&A...510A..18K, 2011MNRAS.412..711T, 2016A&A...586A.103W, Kamp17}. We fitted \xsh\ data of the hydrogen lines \citep{2017A&A...604A..78R} and archival photometry of the dust emission for two sources in M17. The \hii\ region containing the stars and the data are described in Section~\ref{p1:section:M17}. In Section~\ref{p1:section:prodimo} we introduce the model and the modifications applied to the model to enable our analysis. Section~\ref{p1:section:results} shows the results; these are discussed in Section~\ref{p1:section:discussion}. Conclusions are outlined in Section~\ref{p1:section:summary}. 

\section{Description of the data} \label{p1:section:M17}

M17 is one of the most luminous \hii\ regions in the Galaxy 
\citep[$3.6\times10^6$~\Lsun;][]{2009ApJ...696.1278P}. It is located in the Carina-Sagittarius spiral arm at a distance of $1.98\pm0.14$~kpc \citep{2011ApJ...733...25X}. The young cluster NGC\,6618 \citep[$\sim$1\,Myr;][]{1997ApJ...489..698H, 2007ApJS..169..353B, 2009ApJ...696.1278P}, containing 16 O-type stars and over 100 B-type stars, is located at the centre of the \hii\ region \citep[][]{1980A&A....91..186C, 2008ApJ...686..310H}, while the surrounding molecular cloud hosts pre-main-sequence (PMS) stars \citep[][]{1997ApJ...489..698H, 2017A&A...604A..78R}. On the basis of a near-infrared (NIR) excess in the spectral energy distribution (SED) and the presence of hydrogen and CO bandhead emission lines, \citet{1997ApJ...489..698H} identified a sample of massive (6-20~\Msun) young stellar object (YSO) candidates embedded in the molecular cloud. Using X-shooter \citep{2011A&A...536A.105V} mounted on the ESO Very Large Telescope, \citet[hereafter \citetalias{2017A&A...604A..78R}]{2017A&A...604A..78R} carried out a spectroscopic follow-up of 12 of these objects, providing the spectroscopic confirmation that these objects are massive pre-main-sequence stars. The stellar parameters were determined from photospheric line fitting or SED fitting. In the nine cases where the objects have visible photospheres, they modelled the stellar spectrum using the non-LTE radiative transfer code {\sc fastwind} \citep[][]{2005A&A...435..669P, 2012A&A...537A..79R} together with the genetic algorithm {\sc pikaia} \citep[][]{2005A&A...441..711M, 1995ApJS..101..309C}. {Otherwise SED fitting was used to determine stellar properties.} The obtained temperatures, luminosities, and projected rotational velocities allowed them to place these MYSOs in the Hertzprung-Russell Diagram (HRD) and compare their position with pre-main-sequence (PMS) evolutionary tracks from \citet{2009ApJ...691..823H}. Two of the observed objects (B111 and B164) have spectra characteristic of O-type stars and are located at the zero-age main-sequence (ZAMS). 
Two other objects (B215 and B289) show an IR excess longwards of 2.3~$\mu$m but do not show emission lines nor NIR excess in the \xsh\ spectral range. This is a sign that they are surrounded by dusty disks.
Six of the observed objects (B163, B243, B268, B275, B331, and B337) have H\,{\sc i}, O\,{\sc i}, and Ca\,{\sc ii} double-peaked emission lines as well as CO bandhead emission and a NIR excess in their SED; this indicates that they are surrounded by a gaseous disk with a dust component. \citetalias{2017A&A...604A..78R} conclude that they are MYSOs with disks that are probably a remnant of the assembly process on their way to become main-sequence stars with masses between 6 and 15~\Msun\ consistent with having undergone high mass-accretion rates ($M_{\rm acc} \sim 10^{-4} - 10^{-3}$~\Msun\,yr$^{-1}$), hence with formation timescales of $\sim 10^{5}$\,yr.

\subsection{Target objects, observations, data reduction, and contaminants}
\label{sec:targets}

In this paper, we focus on modelling the circumstellar disks of two MYSOs in M17: B243 and B331, whose stellar properties are summarized in Table~\ref{p1:tab:Stellar_parameters} \citepalias{2017A&A...604A..78R}. These two targets show pronounced disk features, namely strong and clearly detected double-peaked hydrogen emission lines up to high order in the Paschen series and an infrared excess. B331 also shows Brackett series lines.  
Their X-shooter spectra cover the optical to NIR (300 - 2500\,nm) with a spectral resolution of 5100 in the UVB arm, 8800 in the VIS arm, and 11\,300 in the NIR arm. The observed spectra are flux calibrated using spectrophotometric standards from the ESO database.

\begin{table*}
 \centering 
 \caption{\normalsize{Stellar properties and extinction parameters of B243 and B331 updated from \citetalias{2017A&A...604A..78R}}.}
\begin{minipage}{\textwidth}
 \centering 
\renewcommand{\arraystretch}{1.4}
\setlength{\tabcolsep}{5pt}
\begin{tabular}{ccclllllll}
 \hline 
 \hline 
Name & Sp. Type & $M$ & \multicolumn{1}{c}{$T_{\rm eff}$} & $\log{g}$ & $v\sin{i}$ & $R_V$ & $A_V$ & $\log{L/L_{\odot}}$ & $R_{\star}$ \\
&  & ${\rm M_\odot}$ & \multicolumn{1}{c}{K} & cm\,$\rm s^{-2}$ & km\,$\rm s^{-1}$ & (SED) & mag & & $R_{\odot}$ \\
\hline 
B243    &      B8 V & 6    &      $13500_{-1250}^{+1350}$ & $4.34_{-0.3}^{\uparrow}$ & $110_{\downarrow}^{+106}$ &$4.7_{-0.8}^{\uparrow}$ &$8.5_{-1.0}^{\uparrow}$ &$3.21_{-0.06}^{+0.07}$ &$7.5_{-0.8}^{+1.0}$ \\ 
B331    &  Late-B & 12 &      $\sim$$13000$ & $-$\footnote{\label{p1:footnote_table1}We assume $\log{g}=4.0$} & $-$\footnote{We assume $v \sin{i} = 110$} &$4.6_{-0.5}^{+0.5}$ &$13.3_{-0.9}^{+0.9}$ &$4.10_{\downarrow}^{+0.37}$\footnote{Value from SED fitting} &$21.8_{-7.2}^{+9.6}$ \\
\hline
\vspace{-15pt}
\end{tabular}
\renewcommand{\footnoterule}{}
\end{minipage}
\label{p1:tab:Stellar_parameters}
\end{table*} 

The H\,{\sc i} circumstellar emission in the spectra of the two sources can be contaminated with various other features, including nebular emission, stellar absorption, circumstellar emission by metallic species, and interstellar absorption. Some of these can and need to be corrected for in order to isolate the hydrogen emission from the disk itself.

The spectrum of B331 shows strong nebular emission, which contaminates the hydrogen circumstellar emission by adding a strong central component. Most of this emission was corrected for using multiple Gaussians fitted to off-source nebular emission as described by \citet{2020A&A...636A..54V}. 
The nebular emission in B243 was not as substantial, and did not require further correction. 
Additionally, the studied hydrogen lines are blended with other emission lines originating from the disk, such as weak Ca\,{\sc ii}\,$\lambda8498,~ \lambda8542,~\lambda8662$\,\AA\ emission in B243, blending the blue wings and peaks of Pa-16, -15, and -13. It is beyond the scope of this study to investigate the emission mechanism of these blending lines. Consequently, we cannot accurately predict their strengths and we refrain from correcting for their presence. 

The H\,{\sc i} cirsumstellar emission lines are superimposed on the stellar hydrogen absorption lines. The measured flux needs to be corrected for this absorption to isolate the disk contribution. The applied procedure to do so is described in section~\ref{p1:subsection:norm_and_corr}. 

\begin{figure*}
    \centering
    \includegraphics[width=0.9\textwidth]{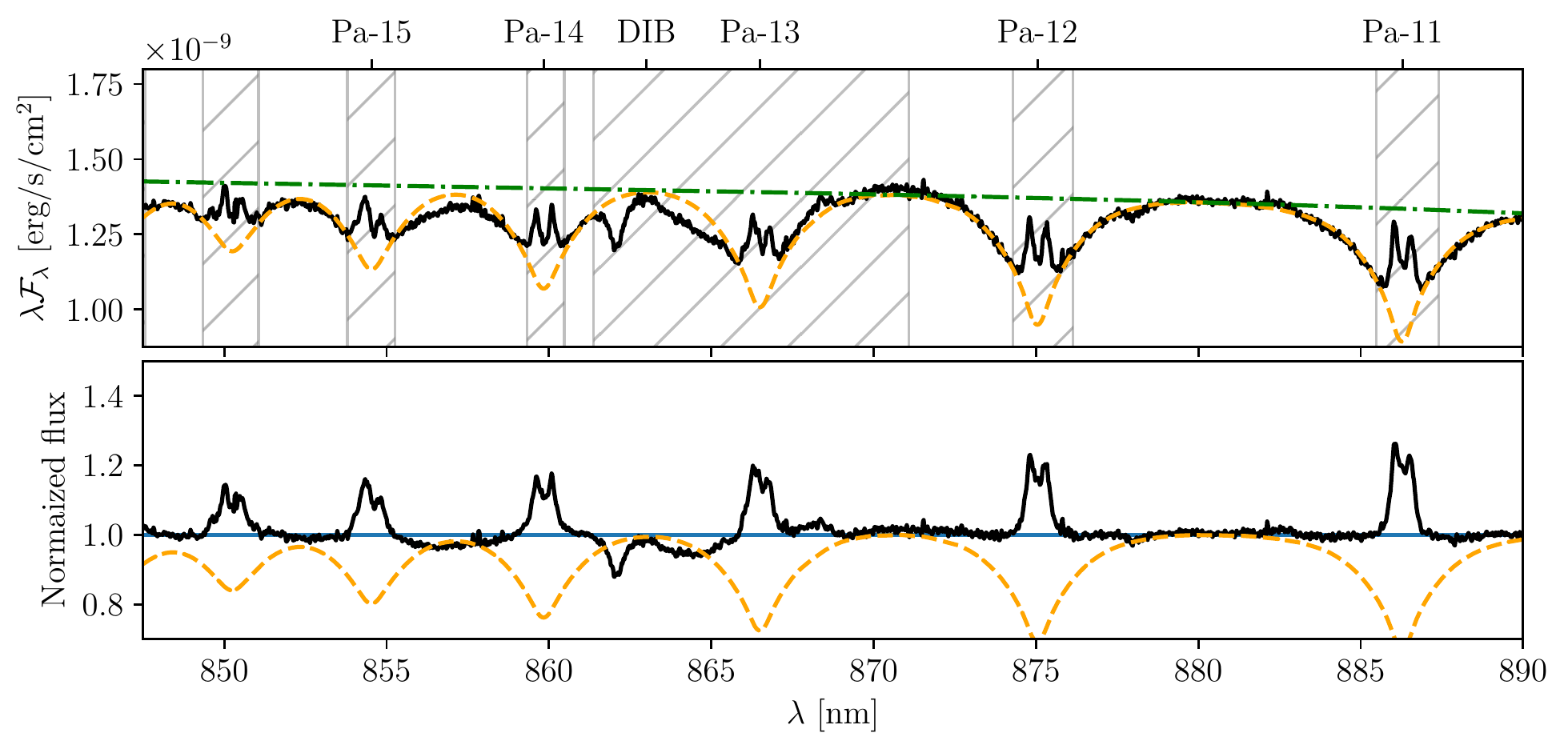}
    \caption{Illustration of the normalization method. {\it Top panel}: The flux calibrated observed spectrum of B243 in black. The photospheric model spectrum is indicated by the orange dashed line and the continuum of this model by the green dash-dotted line. The model spectrum is matched to the observed spectrum using a least squares method in the unhatched wavelength ranges. At the top the location of the Paschen lines and a DIB are indicated. {\it Bottom panel}: The photosphere corrected and normalized observed spectrum  (in solid black) and photospheric model (in dashed blue). The horizontal blue line indicates unity.}
    \label{p1:fig:normalization_example}
\end{figure*}

Finally the spectrum also shows relatively wide absorption features originating from the interstellar medium. These so-called diffuse interstellar bands (DIBs) are likely due to large carbonaceous molecules such as C$_{60}^+$
(e.g. \citeauthor{2006JMoSp.238....1S} \citeyear{2006JMoSp.238....1S}; \citeauthor{2017A&A...606A..76C} \citeyear{2017A&A...606A..76C}; but see \citeauthor{2015Natur.523..322C} \citeyear{2015Natur.523..322C}).
An example of such a feature is seen in Fig.~\ref{p1:fig:normalization_example} at $\lambda$8630\,\AA. We exclude regions with known DIBs when estimating the continuum level. 

Slight variations in the continuum of the normalized spectrum are still visible after applying the correction for stellar and nebular emission. These variations, due to unknown features, irregularities in the response curve, flux calibration, and uncertainties in the correction and normalization method, are the main uncertainty on the data. We conservatively estimate these uncertainties to be approximately 5\% of the continuum on average and assume this to be normally distributed.

\subsection{Photometric data}
The photometric data used in this work is taken from the literature and catalogues. For both stars we included data from DENIS \citep{2000A&AS..141..313F}, 2MASS \citep{2006AJ....131.1163S}, and Spitzer GLIMPSE \citep{2005PASP..117..978R}. WISE photometric data were included for B243 \citep{2010AJ....140.1868W}, as well as points from  \citet{2002AJ....124.1636K} for B331. SOFIA photometry of B331 at 20\,$\mu$m and 37\,$\mu$m \citet{2020ApJ...888...98L} shows very high flux and, due to SOFIA's low spatial resolution, likely contains emission from outer material not belonging to the circumstellar disk. Therefore, it was not included in the analysis. Additionally, the spatial distribution of the SOFIA data does not agree with the 2MASS and Spitzer data, with the SOFIA data being significantly offset with respect to the others.

\subsection{Normalization and correction of stellar features} 
\label{p1:subsection:norm_and_corr}

To correctly model the disk emission lines contaminant stellar absorption features had to be removed, that is the flux absorbed in the stellar atmosphere had to be added back to the observed flux. To do so, a good estimate of the continuum is essential. However, in and around the hydrogen line series it is not possible to accurately make this estimate based on the observed spectra alone, as the Paschen line wings blend to form a 'secondary' continuum flux almost everywhere in the studied part of the spectrum; see the top panel of Fig.~\ref{p1:fig:normalization_example}.
As the stellar properties of B243 and B331 were already determined by \citetalias{2017A&A...604A..78R}, see Table~\ref{p1:tab:Stellar_parameters}, we chose to simultaneously normalize and correct for the stellar features by matching model spectra to the flux calibrated observed spectra, for which we used a BT-NextGen atmospheric model computed with the {\sc phoenix} code \citep{1999ApJ...512..377H, 2012RSPTA.370.2765A}.

The model spectrum, with $T_{\rm eff} = 13000$\,K and $\log g = 4.0$, is rotationally broadened and convolved with the spectral resolution of the observation. Then the flux is scaled to match the observed spectrum using a least squares algorithm only taking wavelength ranges into account free from circumstellar or interstellar features (non-hatched regions in Fig\,\ref{p1:fig:normalization_example}). As the calibrated spectrum has slight flux variations this scaling is done piece wise to find the best agreement between model and observation. 
The thus obtained scaled model is continuum subtracted to obtain the stellar features, which are then added to the observed spectrum. The resulting spectrum now only contains stellar continuum, disk features, and interstellar features. This spectrum is then divided by the stellar continuum from the model to get a normalized spectrum. Fig.~\ref{p1:fig:normalization_example} shows an example of this process in part of the Paschen series regime of B331. A more extended part of the spectrum of both B243 and B331 is shown in Fig.\,\ref{p1:fig:almost_full_spectra}
The uncertainty on the stellar parameters will affect this normalization. Overestimating the surface gravity of the star results in wider hydrogen lines, which would result in stronger wings in the circumstellar emission profile. Underestimating the gravity would result in absorption features in the wings. The uncertainty in the temperature slightly affects the strength of the hydrogen lines. 

\section{Model} \label{p1:section:prodimo}
The double peaked hydrogen emission lines observed in B243 and B331 suggests that the emission originates in a circumstellar disk. Neither of the two objects show strong indication of outflows present in the \xsh\ slit or other images. Therefore, we assume that the bulk of the emission originates from the central star and a circumstellar disk. A disk wind may be present, possibly contributing to the observed hydrogen emission, however, this is not modelled.

The disks are modelled using the radiation thermo-chemical code \prodimo\footnote{Version 1.0 revision 3053} \citep{Woitke09, 2010A&A...510A..18K, 2011MNRAS.412..711T, 2016A&A...586A.103W, Kamp17}. \prodimo\ self-consistently solves the chemistry, gas and dust heating and cooling, as well as line and continuum radiative transfer.
The disk is assumed to be azimuthally symmetric and in steady state, and to follow a Keplerian rotation profile. 

The applied chemical network accounts for 13 elements, namely H, He, C, N, O, Ne, Na, Mg, Si, S, Ar, Fe, and Polycyclic Aromatic Hydrocarbons (PAHs). In the present work, the latter only contribute to the free electron density.  A total of 100 chemical species can form and interact. For the elemental abundances we adopt the Proto-Sun based values of \citet{2003ApJ...591.1220L}. These abundances are not depleted to compensate for the metals that are trapped in dust grains, 
as the chemistry we focus on takes place in the dust free inner region of the disk. The chemical reactions and rate coefficients are taken from the {\sc UMIST\ 2006} database \citep{2007A&A...466.1197W}. 

The dust temperature and continuum emission are solved for the given stellar and interstellar radiation field and dust opacities. The adopted dust composition is a mix of olivine (Mg$_{2}$SiO$_{4}$), amorphous carbon, and vacuum. Dust settling is taken into account following \citet{1995Icar..114..237D} adopting a turbulent viscosity parameters $\alpha = 10^{-3}$. A complete description of the treatment of solids can be found in \citet{Woitke16}.

Next, the gas heating and cooling are solved simultaneously with the chemistry as these are coupled. The gas temperature is based on 100 heating and 93 cooling processes implemented in \prodimo\ revision number 3053. 

 Because the (free) parameters describing the density structure in the disk are found to be most influential on the studied lines (see section \ref{p1:subsection:model_fitting}), we briefly describe their relations. The relevant parameters are the total dust and gas mass $M_{\rm disk}$; inner and outer radius of the disk, $R_{\rm in}$, and $R_{\rm out}$, respectively; radial column density exponent $\epsilon$; reference vertical scale height $H_0$ specified at a reference radius $R_0$, and scale height flaring power index $\beta$. The inner and outer radii specify the extent of the disk. The radial profile of the vertically integrated column density, $\Sigma$, of the disk is given by
\begin{equation}\label{p1:eq:surface_density}
\Sigma(r) = \Sigma_0 \, (r/R_{\rm in})^{\epsilon},
\end{equation} 
where $\Sigma_0$ is determined such that the integration of this profile over the radial extent of the disk results in $M_{\rm disk}$. The vertical density structure at a given radius $r$ is defined by
\begin{eqnarray}
    \rho(z,r) & = & \rho_0(r)\,\exp{\left[ -\left( z/H \right)^{2}/2 \right]},  \\
\end{eqnarray}
and using equation \ref{p1:eq:surface_density}
\begin{eqnarray}
     \Sigma(r) & = & 2 \int_0^\infty \rho(z,r) \,dz,
\end{eqnarray} 
is solved to find the midplane density  $\rho_0(r)$. The vertical scale height $H$ is a function of the distance to the central star, that is 
\begin{equation}
    H(r) = H_0 \left(\frac{r}{R_0}\right)^{\beta},
\end{equation}
where, $H_0$ is the scale height at the reference radius, $R_0$, and $\beta$ is the flaring power. 

The disk is in Keplerian rotation. It is irradiated by a central star of mass $M_{\star}$, effective temperature $T_{\rm eff}$, and luminosity $L_{\star}$. Additionally, an X-ray luminosity is included 
referred to as $L_{\rm X}$ which is plays an important role in the chemistry and thermal balance of the gas. 
A mass accretion rate $\dot{M}$ can be specified as well. This causes viscous heating in the disk. For B243 we include a minor mass accretion, however, the heating resulting from this is minor, $<$1\%, compared to that of the radiation field. The accretion rate was chosen such that the disk would have a reasonable lifetime given its mass. In B331 no effect was seen as result of similar accretion rates, therefore no accretion heating was implemented.

We divide the disk in two regions, referred to as inner disk and outer disk. The inner disk is located closest to the star and is purely gaseous. This zone has a negligible dust mass as temperatures are too high for solid particles to exist for any appreciable amount of time. The hydrogen line emission originates in this part of the disk.  
The outer disk also contains dust, hence determines the infrared continuum emission. The inner radius of the outer disk results from a fit to the SED.
For the outer disk the model parameters are given the label `2'; for instance, the total mass of the outer disk is $M_{\rm 2,disk}$.

\subsection{Model modifications}

\prodimo\ is originally designed to model T\,Tauri and Herbig disks, that is planet forming disks around cooler less-luminous stars, where a photospheric EUV radiation field is not expected to significantly affect disk processes. Therefore, hydrogen-ionizing photons with energies between 13.6 and 100\,eV were previously ignored. For hotter stars the EUV contribution becomes stronger, and needs to be considered. 
\prodimo\ does include X-ray radiation \citep{2011A&A...526A.163A, 2018A&A...609A..91R}, defined as photons with energies > 100 eV, as the ionizing properties of such photons are essential for the chemistry in the disk  \citep{1997ApJ...480..344G}. In order to properly treat hydrogen ionization, we extended the code to also include the EUV radiation field.

The standard atomic model for hydrogen in \prodimo\ considers 25 $(n,\ell)$\footnote{With $n$ the principal quantum number, indicating the excitation level, and $\ell$ the azimuthal quantum number.} energy levels up to $n$\,=\,5. 
This does not allow for a prediction of the observed Paschen and Brackett series, reaching up to $n$\,=\,16. Therefore, the model atom was replaced by a new model using the NIST atomic database \citep{2019APS..DMPN09004K} to include the energy levels up to $n$\,=\,20 and the transitions between these states. The collisional excitation cross sections are calculated using \citet{1968slf..book.....J} as
\begin{equation}
    q_{ul} = 2.16 \left(\frac{E_0}{kT}\right)^{-1.68} T^{-3/2} f_{lu} \frac{g_l}{g_u} ~~~{\rm [cm}^3 {\rm s}^{-1}{\rm ],}
\end{equation}
with $E_0$ the energy of the transition between energy levels $u$ and $l$, $k$ the Boltzmann constant, $T$ the temperature of the gas, $f_{lu}$ the oscillator strength of the transition, and $g_l$ and $g_u$ the statistical weights of the lower and upper levels. 

For the treatment of photo-ionization, see for example \citet{1978stat.book.....M}, we use
\begin{equation}
    \alpha^{\rm bf}_\nu = 7.91 \times 10^{-18} \frac{n}{Z^2} g_\textsc{ii}(\nu, n) \left(\frac{\nu_0(n)}{\nu}\right)^3 ~~~ {\rm [cm}^2{\rm ]}, 
\end{equation}
with $\nu$ the frequency of a given photon, the charge of the ion $Z=1$ for hydrogen, $g_\textsc{ii}$ the  bound-free Gaunt factor, which we assume to be 1, and $\nu_0(n)$ the ionization frequency for energy level $n$. 

\subsection{Model fitting and model grid}\label{p1:subsection:model_fitting}

\begin{figure}[b!]
    \centering
    \includegraphics[width=\columnwidth]{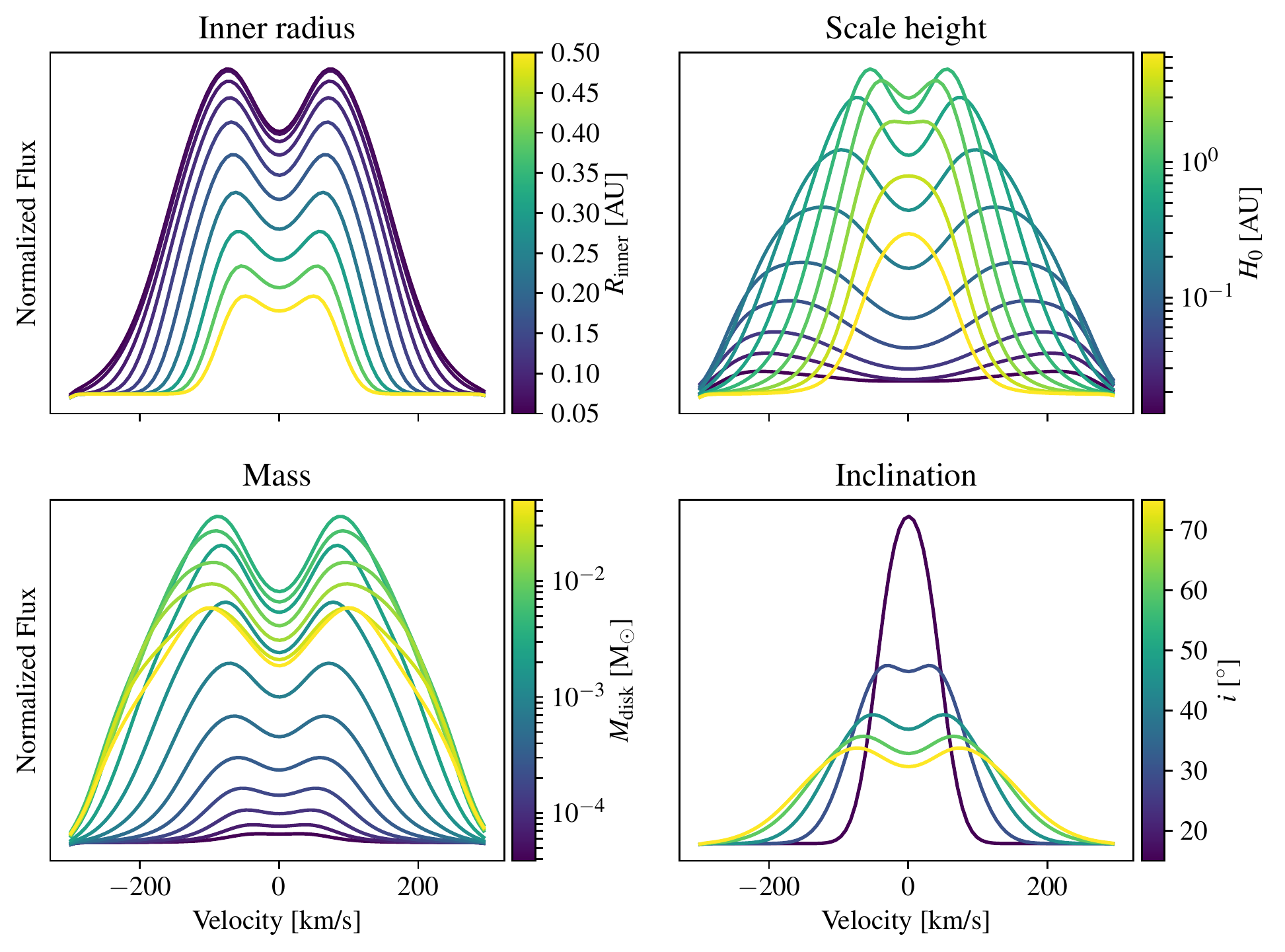}
    \caption{Line profile of Pa-14 from the model of B243 for various parameter values. In each of the panels one parameter is varied, while the others remain fixed at an inner radius of 0.06~AU, a reference scale height of 0.5~AU, a total disk mass of $10^{-3}\ {\rm M}_\odot$, and an inclination of 75$^\circ$.}
    \label{p1:fig:parameter_effect}
\end{figure}

We fitted the emission lines and SED separately, by varying certain parameters of the inner and outer disk, respectively. 
In the inner disk the hydrogen line profiles were fitted using an interpolated 4-dimensional grid of models. The free parameters of this grid are inner radius ($R_{\rm in}$), reference scale height ($H_0$), inner disk mass ($M_{\rm inner}$), and inclination ($i$). These parameters have a strong effect on the resulting line profiles. This is illustrated in Fig.~\ref{p1:fig:parameter_effect}, where large variations in line strength and shape are visible. Increasing the inner radius of the disk results in weaker and narrower lines, as the disk moves further from the star where it is less illuminated and orbital velocities are lower. The smallest sampled scale heights result in weaker and wider lines. As the material is confined closer to the mid-plane of the disk, less of the stellar radiation reaches the disk; moreover, it is harder for this radiation to reach more (radially) distant parts of the disk due to the high density. Consequently, line emission emanates from regions with relatively high Keplerian velocities. As the volume density of the disk decreases with increasing scale height and light reaches larger radial distances, the peak separation decreases. The line strength initially increases, because more radiation is intercepted by the disk.  However, at the highest scale heights, the decrease in volume density starts to outweigh the increased illumination and the line strength decreases again. Increasing the mass of the disk results in a higher density and thus a stronger emission line, until it saturates from which point on the line strength will decrease. Changing the inclination angle changes the projected velocity distribution of the emission. 

Other parameters such as the flaring power, $\beta$, and column density exponent, $\epsilon$, are degenerate with the varied parameters in how they influence the line profiles. This degeneracy results from the small area of the disk (see section \ref{p1:subsection:physical_properties}) from which the H\,{\sc i} emission lines originate. Therefore, the surface density exponent and disk mass affect the density of the emission region in a similar way.
The flaring power affects the scale height of this region in a similar way as the reference scale height. A fiducial flaring power (of e.g. 1.2) results in an extremely large vertical extent in the outer parts of the inner disk as a relatively large scale height is required close to the star. To prevent this, we adopt a value of $\beta=0.5$ for the inner disk. 

\begin{figure}[b!]
    \centering
    \includegraphics[width=\columnwidth]{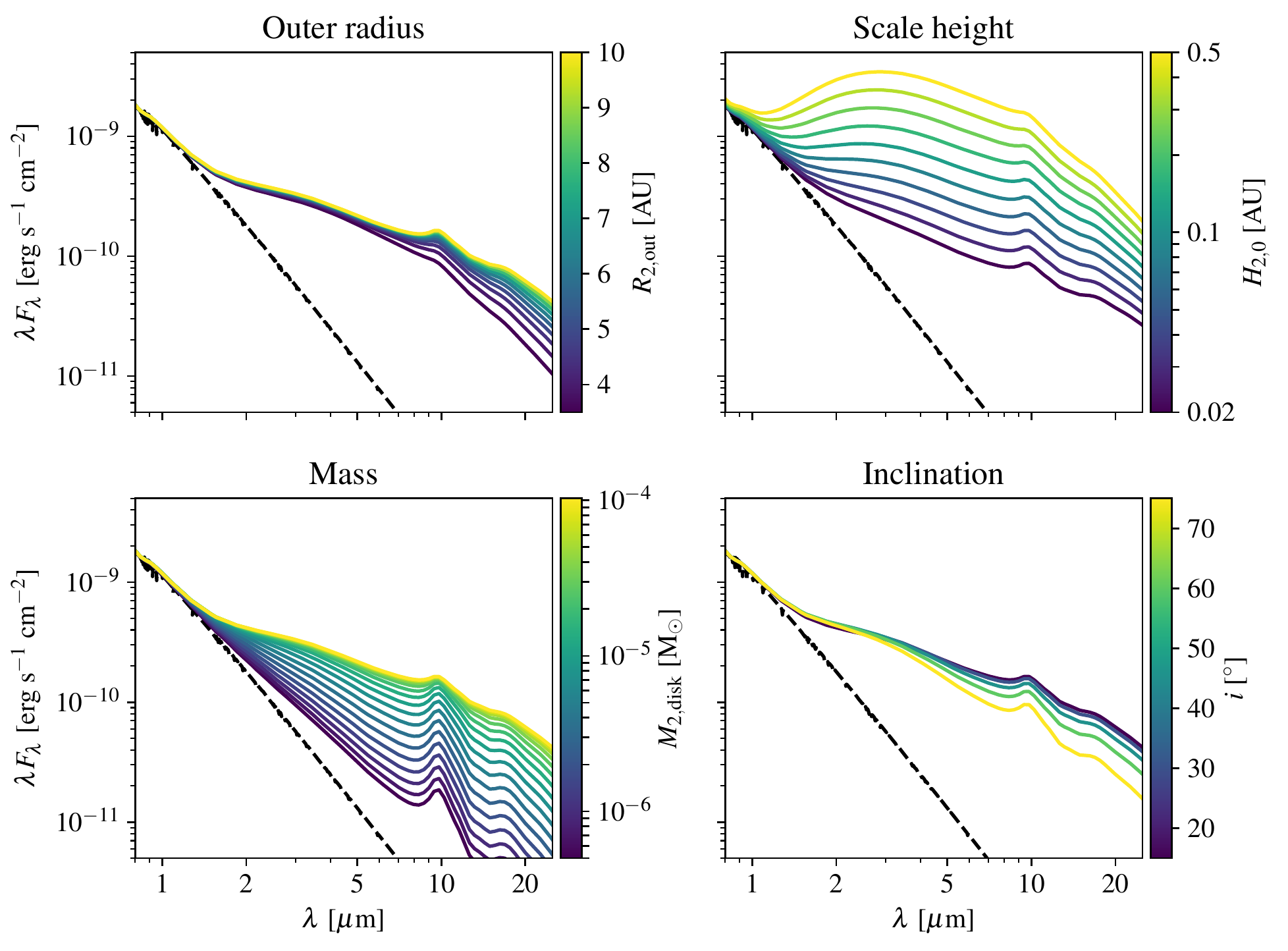}
    \caption{Models of the SED for B243 for various parameter values. In each of the panels one parameter is varied, while the others are held fixed at an outer radius of 10\,AU, a reference scale height of 0.04\,AU, a disk mass of 10$^{-4}$\,M$_\odot$, and an inclination of 15$^\circ$. }
    \label{p1:fig:parameter_effect_SED}
\end{figure}

The outer disk is modelled using a similar grid, varying the outer radius, scale height, and mass. An example of the effects of the parameters on the SED is shown in Fig.~\ref{p1:fig:parameter_effect_SED}. The inclination has a modest effect on the SED. For consistency with the inner gaseous disk we set this parameter to the best fit value of the line profiles. The scale height has a strong effect on the continuum disk emission. For the relatively massive disk shown here, a larger outer radius allows for cooler dust causing stronger mid-infrared emission. However, for a relatively low disk mass (not shown here) the redistribution of mass that results from assuming a larger outer radius causes the inner regions of the disk to become optically thin, reducing the near-infrared flux.  The mass of the disk affects the full spectral range. For a higher mass the emission increases until the disk becomes optically thick. This happens first at shorter wavelengths. 

The parameter values of the calculated grid points are listed in Tab.~\ref{p1:tab:parameter_ranges}. The range of adopted values allow for both much weaker and much stronger lines than observed. As the range can cover several orders of magnitude we opt for a relatively coarse grid and interpolate between the models using an N-dimensional linear interpolation algorithm \citep{2020SciPy-NMeth}. With this interpolation a significantly higher resolution grid consisting of 3\,125\,000 points is calculated. This finer grid is used to fit the data. The interpolated data is verified to be accurate by calculating full models at interpolated grid points. Typically, the interpolated line profiles diverge by $<$1\% from full model calculations, however in some cases the stronger lines can differ by up to 10\%.

A summary of input parameters for \prodimo\ is shown in Tab.\,\ref{p1:tab:parameters}. 
Parameters not listed are identical to the \prodimo\ standard parameter values obtained in the {\sc diana} project \citep{Woitke16}. 
Full input files can be found online\footnote{A reproduction package can be found at \href{https://doi.org/10.5281/zenodo.7024846}{https://doi.org/10.5281/zenodo.7024846} \label{p1:fn:Zenodo}}. Parameter values listed in the table that are held fixed are the result of exploratory fitting. The fixed parameters are either degenerate with other parameters (see section \ref{p1:subsection:model_fitting}) or do not impact the fitting strongly, therefore should not be considered well constrained. Notably, the outer radius of the inner disk is poorly determined as the H\,{\sc i} lines form within the first one to two AU of the inner disk, see Fig.~\ref{p1:fig:B243_physical_properties} and Fig.~\ref{p1:fig:B331_physical_properties}. 

\begin{table}[t!]
\begin{threeparttable}
\caption{Overview of the model grid used to analyse B243 and B331. }
\label{p1:tab:parameter_ranges}
\small
\begin{tabular}{l|l||l|l|l|l}
\rowcolor{lightgray} \multicolumn{6}{c}{Inner disk}                                                                                                   \\ \hline \hline
                                           &         & $R_{\rm in}$ {[}AU{]} & $H_0$ {[}AU{]} & $M$ {[}M$_\odot${]}             & $i$ {[}$^\circ${]}  \\ \hline
\multicolumn{1}{l|}{\multirow{3}{*}{B243}} & min     & 0.05                  & 0.0139         & $3.87\times 10^{-5}$            & 15                  \\
\multicolumn{1}{l|}{}                      & max     & 0.5                   & 6.46           & 0.05                            & 75                  \\
\multicolumn{1}{l|}{}                      & steps   & 10                    & 13             & 15                              & 5                   \\ \hline
\multicolumn{1}{l|}{\multirow{3}{*}{B331}} & min     & 0.11                  & 0.1            & $1.94\times 10^{-4}$            & 15                  \\
\multicolumn{1}{l|}{}                      & max     & 0.30                  & 10.0           & 0.25                            & 75                  \\
\multicolumn{1}{l|}{}                      & steps   & 10                    & 10             & 15                              & 5                   \\ \hline
\multicolumn{2}{l||}{Interpolation}            & 50                    & 50             & 50                              & 25                  \\ \hline
\multicolumn{2}{l||}{Spacing}                        & Log                   & Log            & Log                             & Lin                 \\ \hline
\end{tabular}
\begin{tabular}{l|l||l|l|l}
\rowcolor{lightgray} \multicolumn{5}{c}{Outer disk}                                                                                                   \\ \hline \hline
                                           &         & $R_{\rm 2,out}$ {[}AU{]} & $H_{2,0}$ {[}AU{]} & $M_2$ {[}M$_\odot${]}    \\ \hline
\multicolumn{1}{l|}{\multirow{3}{*}{B243}} & min     & 3.5                   & 0.02           & $5.0 \times 10^{-7}$            \\ 
\multicolumn{1}{l|}{}                      & max     & 10                    & 0.5            & $1.03\times10^{-4}$             \\ 
\multicolumn{1}{l|}{}                      & steps   & 10                    & 10             & 17                              \\ \hline 
\multicolumn{1}{l|}{\multirow{3}{*}{B331}} & min     & 110                   & 0.05           & $10^{-3}$                       \\ 
\multicolumn{1}{l|}{}                      & max     & 500                   & 10.0           & 0.5                             \\ 
\multicolumn{1}{l|}{}                      & steps   & 10                    & 10             & 10                              \\ \hline 
\multicolumn{2}{l||}{Interpolation}        & 50      & 50                    & 50             \\ \hline 
\multicolumn{2}{l||}{Spacing}              & Log     & Log                   & Log            \\ \hline 

\end{tabular}
\begin{tablenotes}
 \item The numbers indicate the minimum and maximum values of a parameter, and the number of steps between those values. Interpolation and spacing indicate the number of grid points and spacing between steps in the interpolated grid. The inner disk grid for B243 consists of 9750 models and for B331 of 7500 models.
\end{tablenotes}
\normalsize
\end{threeparttable}
\end{table}

The hydrogen emission lines studied in this work are for B243: 
Pa-$\beta$ through P-16, save for Pa-8 and Pa-10 which are in regions of strong telluric absorption. 
The same lines are studied for B331 as well as Br-$\gamma$, Br-10, Br-11 Br-12, Br-14, and Br-16.
The disk parameters are determined by fitting a subset of the lines simultaneously. The subset aims to include the lines most representative of the inner region of the disk. This subset consists of Pa-16 to Pa-9 for B243 and for B331 Br-16 to Br-11 is added to that set. Additionally, each line if fitted separately allowing the inner radius, scale height, and mass to vary. The inclination is fixed to the best fit value of the combination of lines. 

\begin{table*}
\centering
\caption{Overview of the input parameters for both the central star and the disk.}
\begin{threeparttable}
\small
\begin{tabular}{c|l|l|l|l}
\hline \\[-9pt] 
Component                           & Name                      & Symbol                & B243 Value        & B331 Value      \\  \hline \\[-8pt]
\multirow{4}{*}{Central star} 
& Mass                      & $M_\star$             & 6.0 M$_\odot$     & 12 M$_\odot$      \\
& Effective temperature     & $T_{\rm eff}$         & 13\,500 K           & 13\,000 K           \\
& Luminosity                & $L_\star$             & 1\,622 L$_\odot$    & 12\,500 L$_\odot$   \\
& X-ray luminosity\tnote    & $L_{\rm X}$           & $10^{32}$ erg\,s$^{-1}$ & $5 \times 10^{32}$ erg\,s$^{-1}$  \\
& Accretion rate\tnote      & $\dot{M}$             & $10^{-7}$ M$_\odot\,{\rm yr^{-1}}$ & - \\ \hline \\[-8pt]

\multirow{7}{*}{\shortstack{Density structure \\ inner disk}}
& Mass                      & $M_{\rm disk}$       & Varied    & Varied        \\
& Inner radius              & $R_{\rm in}$         & Varied    & Varied        \\
& Outer radius			    & $R_{\rm out}$		& $> 2$\,AU (set at 3.5 AU)   & $> 2$\,AU (set at 20 AU)         \\
& Column density exponent   & $\epsilon$           & 0.0       & -1.0          \\
& Reference scale height    & $H_{0}$              & Varied    & Varied    	\\
& Reference radius		    & $R_{0}$              & 3.5 AU    & 10 AU         \\
& Flaring power-index       & $\beta$				& 0.5       & 0.5			\\ \hline \\[-8pt]

\multirow{7}{*}{\shortstack{Density structure \\ outer disk}}
& Mass                      & $M_{\rm 2,disk}$        & Varied          & Varied              \\
& Inner radius              & $R_{\rm 2,in}$          & 3.0 AU          & 100 AU              \\
& Outer radius			    & $R_{\rm 2,out}$			& Varied          & Varied              \\
& Column density exponent   & $\epsilon_2$            & -1.0            & -1.5                 \\
& Reference scale height    & $H_{2,0}$               & Varied          & Varied    	        \\
& Reference radius		    & $R_{2,0}$               & 3.5 AU          & 100 AU         	    \\
& Flaring power-index       & $\beta_2$ 				& 1.2             & 1.2					\\ \hline \\[-8pt]

\multirow{5}{*}{Dust in outer disk}
& Minimum grain size        & $a_{\rm min}$                     & \multicolumn{2}{c}{0.5\,$\mu$m}           \\
& Maximum grain size        & $a_{\rm max}$                     & \multicolumn{2}{c}{1000\,$\mu$m}          \\
& Grain size exponent       & $a_{\rm pow}$                     & \multicolumn{2}{c}{3.5}                   \\
& Turbulent mixing parameter& $\alpha_{\rm settle}$             & \multicolumn{2}{c}{$10^{-3}$}               \\ 
& Dust-to-gas ratio         & $M_{\rm dust} / M_{\rm gas}$      & \multicolumn{2}{c}{0.01}                 \\ \hline \\[-8pt]

\multirow{2}{*}{Other}
& Distance                  & $d$                               & \multicolumn{2}{c}{2.0\,kpc}              \\
& Inclination               & $i$                               & \multicolumn{2}{c}{Varied}                
\end{tabular}
\normalsize
\end{threeparttable}
\label{p1:tab:parameters}
\end{table*}

\section{Results} \label{p1:section:results}

We present the results of our fitting efforts of the photometric data and X-shooter spectra using \prodimo. First, we discuss the source of the hydrogen ionization in section~\ref{p1:subsection:source_of_ionization}. Subsequently, we describe the fitting results for B243 and B331 in sections \ref{p1:subsection:results_B243} and \ref{p1:subsection:results_B331}. Correlations between parameters are briefly presented in appendix~\ref{p1:subsection:correlations}. Finally, we present the derived physical properties of the modelled disks in section~\ref{p1:subsection:physical_properties}. All results are also available online$^{\ref{p1:fn:Zenodo}}$.

Only the disk contribution to the emission is fitted. For the stellar contribution we use the best fit properties as listed in Tab.~\ref{p1:tab:Stellar_parameters}. We report the scale heights at the inner radius of the disk rather than at the reference radius, which is used to set up the grid of models. The mass is translated to the column density at the inner radius of the disk, as the latter is better constrained. These quantities are calculated using the equations in Section~\ref{p1:section:prodimo}. An overview of the best fit parameters is shown in Tab.~\ref{p1:tab:fit_results}.

\subsection{Source of hydrogen line emission}\label{p1:subsection:source_of_ionization}
The spectra of both B243 and B331 show significant hydrogen line emission. 
The mechanism through which these lines form is recombination, as the temperatures in these disks are not high enough to allow collisional excitation of the relevant levels. 
Photo-ionization by the central stars cannot account for the hydrogen ionization rates (see section \ref{p1:subsection:physical_properties} and Figs.~\ref{p1:fig:B243_physical_properties} and \ref{p1:fig:B331_physical_properties}) required to produce the observed line strengths -- the stars are simply too cool ($T_{\rm eff}$ being 13\,500\,K and 13\,000\,K). Therefore, an alternative source of ionization in the inner part of the disk is required. In the simulations we find charge exchange reactions between ionized sulfur (S$^{+}$) and hydrogen, that is
\begin{eqnarray}
 {\rm S^{+}} + {\rm H} \rightarrow {\rm S} + {\rm H^{+}}, 
\end{eqnarray}
to be the dominant source of hydrogen ionization.
Sulfur is ionized by UV photons with energies of at least 10.3 eV and then collides with neutral hydrogen, ionizing it. Other charge exchange reactions with hydrogen also take place, however these do not contribute to the total hydrogen ionization significantly. 

\begin{table}[]
\centering
\begin{threeparttable}[width=\columnwidth]
\renewcommand{\arraystretch}{1.35}
\caption{Table with the best fit model parameters and their 1$\sigma$ uncertainties. \label{p1:tab:fit_results}}
\begin{tabular}{l|l|l}
\hline \hline
Parameter                                & \multicolumn{1}{c}{B243}               & \multicolumn{1}{|c}{B331}              \\ \hline
\rowcolor{lightgray} \multicolumn{3}{c}{Inner disk}                                                                        \\\hline
$R_{\rm in}$ {[}R$_\star${]}             & $1.4_{\downarrow}^{+0.4}$              & $1.1_{\downarrow}^{\rm ND}$            \\
$H_0$ {[}AU{]}                           & $0.527_{-0.06}^{\rm ND}$               & $2.4_{\rm ND}^{\rm ND}$                \\
$M_{\rm disk}$ {[}M$_\odot${]}           & $5_{\rm ND}^{\rm ND} \times 10^{-4}$   & $1.8_{\rm ND}^{\rm ND} \times 10^{-4}$ \\
$i$ {[}$^\circ${]}                       & $75_{-10}^{\uparrow}$                  & $75_{\rm ND}^{\uparrow}$               \\
$H_{\rm inner}$ {[}R$_\star${]}          & $1.8_{-0.2}^{+0.3}$                    & $2.5_{\rm ND}^{\rm ND}$                \\
$\Sigma_{\rm inner}$ {[}g cm$^{-2}${]}   & $124_{\rm ND}^{\rm ND}$                & $1.1_{\rm ND}^{\rm ND}\times 10^3$     \\ \hline
\rowcolor{lightgray} \multicolumn{3}{c}{Outer disk}                                                                        \\ \hline
$R_{\rm 2,out}$ {[}AU{]}                  & $6.5_{-2.4}^{+2.1}$                   & $246_{\downarrow}^{\uparrow}$               \\ 
$H_{2,0}$ {[}AU{]}                       & $0.039_{-0.005}^{+0.003}$              & $0.18_{-0.02}^{+0.02}$                 \\
$M_{\rm 2,disk}$ {[}M$_\odot${]}         & $1.0_{-0.3}^{\uparrow} \times 10^{-4}$ & $0.44_{-0.42}^{\uparrow}$              \\
$H_{\rm 2,inner}$ {[}$R_\star${]}        & $0.9_{-0.1}^{+0.1}$                    & $1.8_{-0.2}^{+0.2}$                    \\
$\Sigma_{\rm 2,inner}$ {[}g cm$^{-2}${]} & $14_{-5}^{+27}$                        & $55_{-36}^{\uparrow}$                \\ \hline
\end{tabular}
\begin{tablenotes}[flushleft]
\small
\item The inner disk properties are determined by fitting a subset of hydrogen lines simultaneously. An arrow indicates that the confidence interval extends beyond the model grid. ND indicates that the uncertainty could not be determined as it is smaller than the step size of the grid.
\end{tablenotes}
\end{threeparttable}
\end{table}

\begin{figure*}
    \centering
    \includegraphics[width=0.9\textwidth]{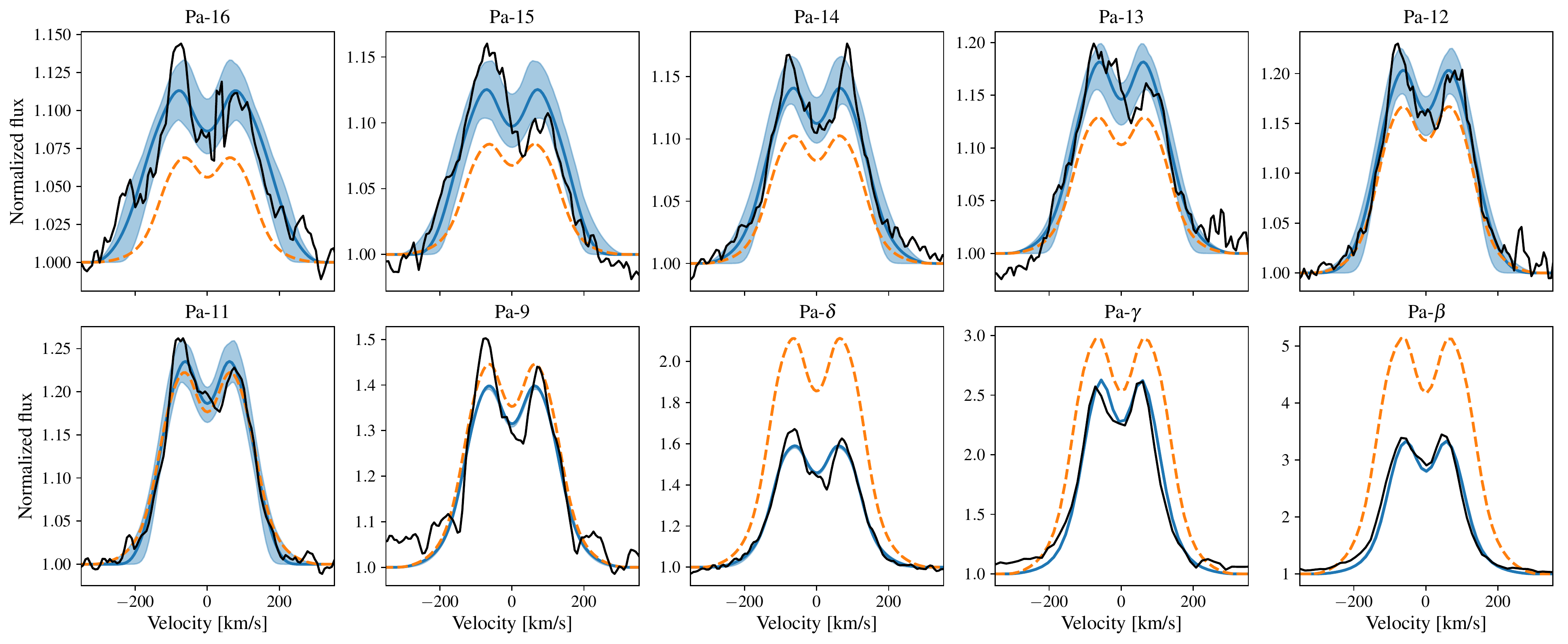}
    \caption{Best fitting models in blue to the observed lines in black for B243. The blue shaded regions indicate models that fall within the 1\,$\sigma$ confidence interval. Each line is fitted individually with the inclination set to $75^\circ$, the respective parameters are shown in Fig.~\ref{p1:fig:B243_B331_parameter_trend}. The orange profiles indicate the best fit when fitting Pa-16 to Pa-9 simultaneously. Notice the factor $\sim$30 range in line strengths covered by the Paschen series.
    \label{p1:fig:B243_individual_line_fits}}
\end{figure*}

\begin{figure}
    \centering
    \includegraphics[width=\columnwidth]{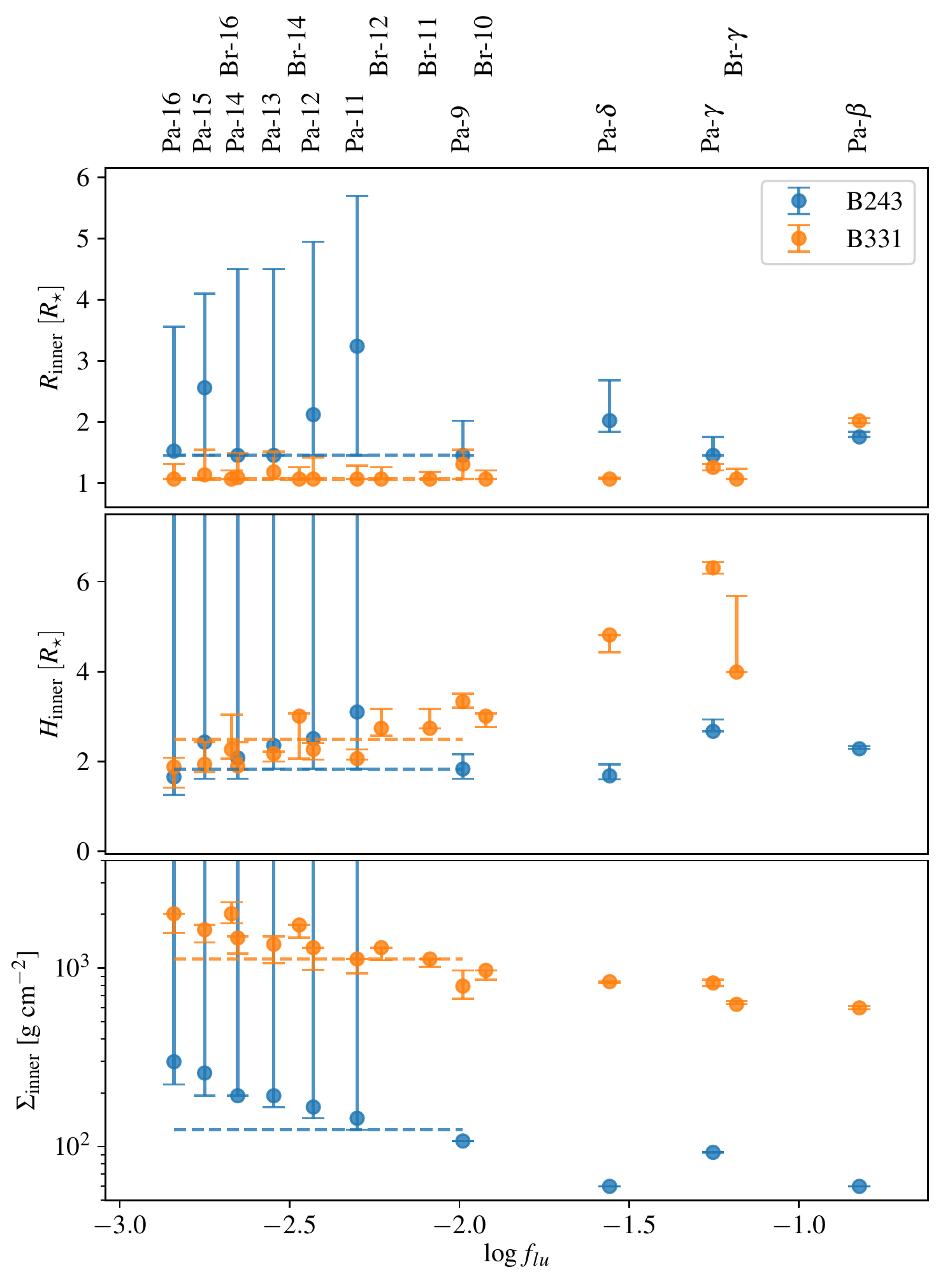}
    \caption{Parameter values of the inner disk with their uncertainties for each modelled line of B243 and B331 as function of their oscillator strength. The horizontal dashed lines indicate the best fit parameters for B243 when fitting Pa-9 and Pa-11 to Pa-16 simultaneously and for B331 when fitting Pa-9, Pa-11 to Pa-16 and Br-11 to Br-16. The corresponding line profiles are shown in Fig.~\ref{p1:fig:B243_individual_line_fits} and Fig.~\ref{p1:fig:B331_individual_line_fits} for B243 and B331, respectively. For each fit an inclination of 75$^\circ$ is used. 
    \label{p1:fig:B243_B331_parameter_trend}}
\end{figure}

\begin{figure}
    \centering
    \includegraphics[width=\columnwidth]{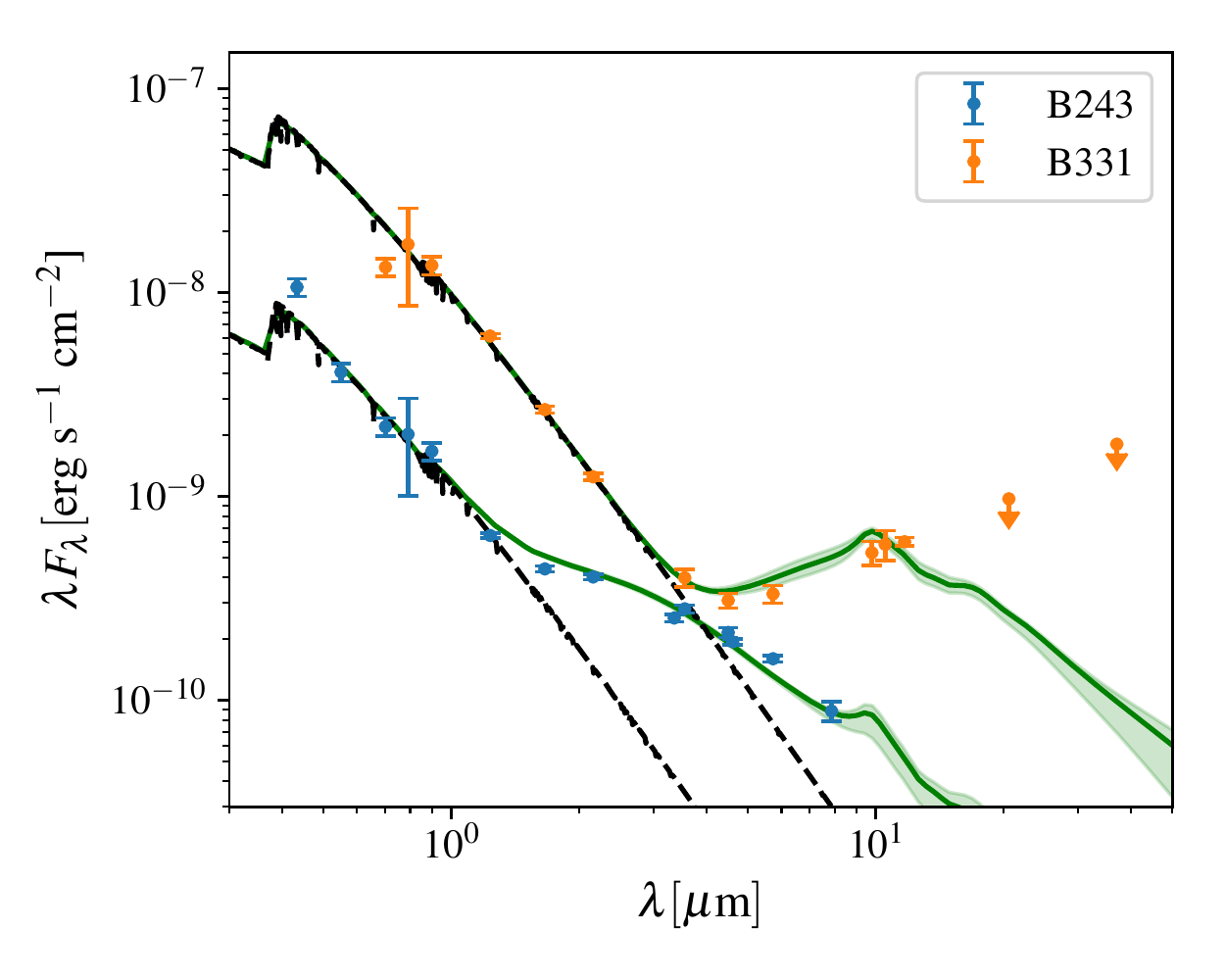}
    \caption{Observed and best fit SED of B243 and B331. For each star, the scatter points indicate the dereddened observed photometry and the black dashed lines the Kurucz model used as input for the stellar spectrum in \prodimo. The best fitting models are drawn in green, with the shaded region indicating the 1$\sigma$ uncertainty. For B331 the photometry from \citet{2020ApJ...888...98L} is indicated as upper limits; these are not considered in the fitting procedure. }
    \label{p1:fig:SED_fit}
\end{figure}

\subsection{B243} \label{p1:subsection:results_B243}

\subsubsection{Hydrogen emission lines}

All of the Paschen lines in B243 show clear double peaked emission consistent with being formed in a rotating disk; see Fig.~\ref{p1:fig:B243_individual_line_fits}. Higher Pa-series lines become weaker and display more pronounced wings and a larger velocity separation of the peaks, see Fig.~\ref{p1:fig:disk_velocity} and \citetalias{2017A&A...604A..78R}. The range in line strength is large, Pa-$\beta$ peaking about three times over continuum while for Pa-16 this is only about 10\%. 

The blue lines in Fig.\,\ref{p1:fig:B243_individual_line_fits} show fits to each line individually. These match the observed profiles very well, save for the blue wing and blue peak of Paschen 16, 15, and 13, which are contaminated with superimposed Ca\,{\sc ii} triplet emission (see Section~\ref{sec:targets}). 
The corresponding model parameters are shown in orange in Fig.~\ref{p1:fig:B243_B331_parameter_trend} as function of oscillator strength (bottom axis) and line identifier (top axis). The parameters derived from the Paschen lines show fair agreement with one another though we note that for this star the uncertainties are sizable, in part due to some level of degeneracy in several of the model parameters. For instance, the inner radius $R_{\rm in}$ and scale height $H$ both impact the line broadening -- increasing $R_{\rm in}$ leads to narrower lines requiring a lower scale height and higher mass (see Fig.\,\ref{p1:fig:parameter_effect}). We find that the Paschen lines indicate a small inner radius, that is the disk extends to the stellar surface to within the order of a stellar radius ($R_\star$ = 0.035\,AU). The surface density at the inner radius is at least $\sim$100\,g\,cm$^{-2}$. 
The vertical scale height of the disk at its inner radius is $\gtrsim$ 2\,$R_{\star}$, which is significantly larger than implied by vertical hydrostatic equilibrium. Such a puffed-up inner zone is quite common in HAeBe stars \citep[e.g.][]{2001ApJ...560..957D} and may originate from its direct exposure to stellar light (see also Section \ref{p1:sec:scale_height_discussion}). The radial extent of the inner disk cannot be constrained from the hydrogen lines, hence, such constraints -- if any -- should come from the dust modelling.

The horizontal dashed line in Fig.~\ref{p1:fig:B243_B331_parameter_trend} shows the best fit parameters when analysing Pa-9 and Pa-11 to Pa-16 simultaneously. The corresponding line profiles are shown in orange in Fig.~\ref{p1:fig:B243_individual_line_fits}. We find a best fit inclination of 75$^\circ$, which is used for the individual line fits. The best value for the column density at the inner radius of the disk lies lower than the confidence intervals of some of the individual fits. This is due to a local minimum in the $\chi^2$. The combined fit recovers all line strengths -- that overall vary by a factor of $\sim$30 -- to within a factor of two. It does reveal a systematic trend in that the model under-predicts the higher (i.e. weaker) Paschen lines and over-predicts the lower Paschen (i.e. stronger) lines.
We discuss this discrepancy in Section~\ref{p1:section:discussion}.

\subsubsection{SED}

The NIR excess of B243 is well fitted with dust particles emitting at an almost constant temperature of $\sim$1500\,K; see Fig.~\ref{p1:fig:SED_fit}. This temperature is about the condensation temperature of silicate-based and carbonaceous grains placing the inner radius of the dust disk at 3\,AU distance from the star. We use the best fit inclination of the inner gaseous disk of 75$^\circ$. We find that the outer radius of the disk is poorly constrained; the best fit lies at $\sim$6.5~AU from the central star. The disk may appear truncated due to some self-shadowing. Longer wavelength photometry is needed to better investigate the extend of the disk.
 A very small truncated dust disk, containing only a small amount of hot dust. The scale height at the inner rim of the dusty outer disk is 0.9$^{+0.1}_{-0.1}\,R_\star$, therefore 
the dust is more confined to the disk mid-plane than is the gaseous material at the inner edge of the inner disk, consistent with the inner most parts of the disk being puffed up. 

We find a lower limit of the surface density at this inner rim of the dusty disk of $\sim$9\,g\,cm$^{-2}$, the material becoming optically thick and insensitive to further increases of the surface density. 
\\

In summary, the gas and dust analysis points to B243 having a disk with a best fit outer radius of 6.5 AU of which the inner part, up to 3 AU, is dust free. The mass contained in this small disk is at least $\sim 2 \times 10^{-5}$\,\Msun\ when adopting a surface density exponent $\epsilon = -1$ for both the inner and outer disk, or $\sim3 \times 10^{-4}$\,\Msun\ when assuming $\epsilon=0$ for the inner disk.
As we lack photometric and spectroscopic information at wavelengths longwards of 10\,$\mu$m we cannot exclude the presence of a (tenuous) cold disk further out, but a gap or self-shadowed region is should be present based on the best fit model.

\subsection{B331} \label{p1:subsection:results_B331}

\begin{figure*}
    \centering
    \includegraphics[width=0.72\textwidth]{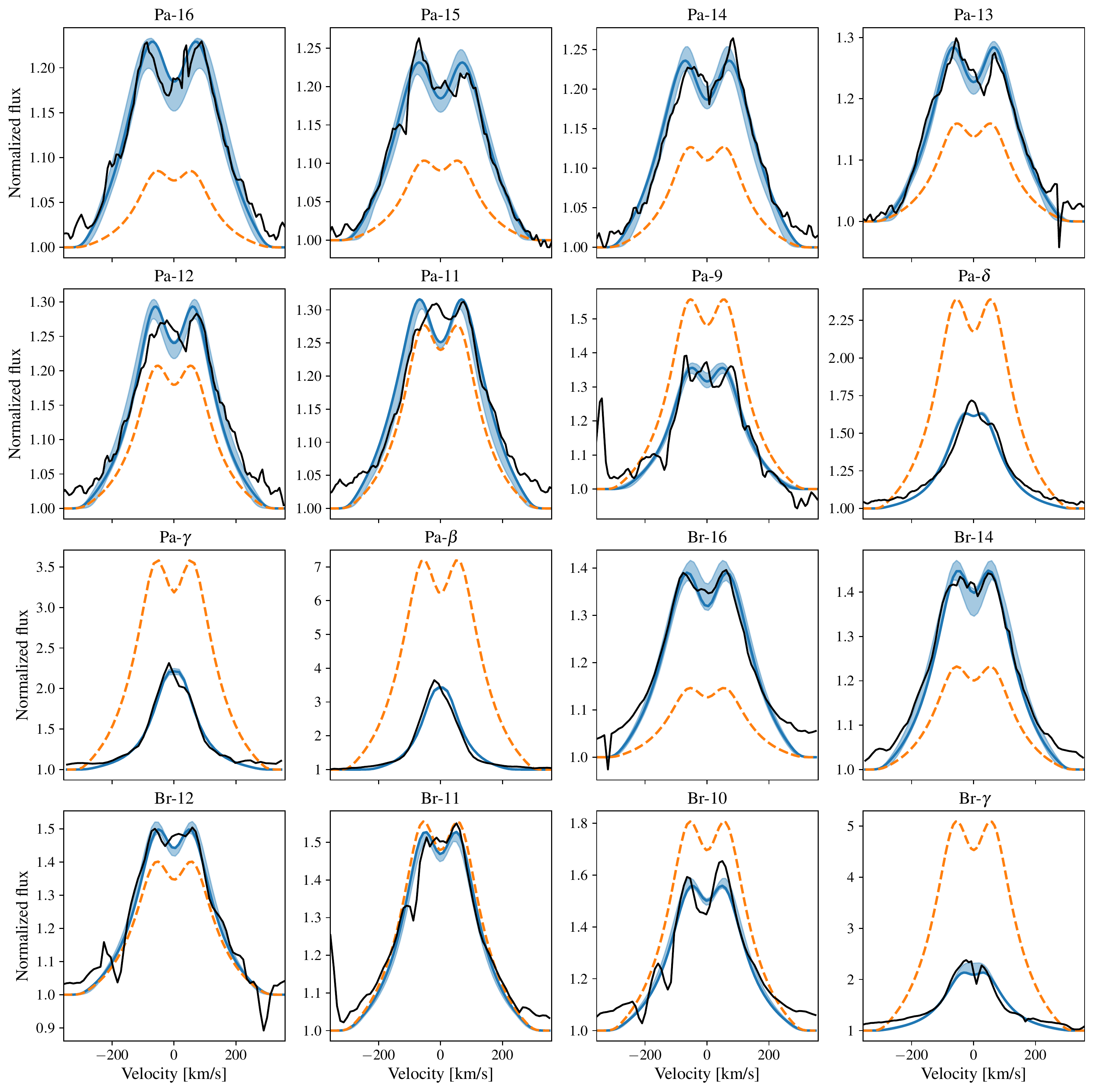}
    \caption{Best fitting models in blue to the observed lines in black for B331. The blue shaded regions indicate models that fall within the 1\,$\sigma$ confidence interval. Each line is fitted individually with the inclination fixed at 75$^\circ$, the respective parameters are shown in Fig.~\ref{p1:fig:B243_B331_parameter_trend}. The orange profiles indicate the best fit when fitting Pa-16 to Pa-9 and Br-16 to Br-11 simultaneously. Notice the factor $\sim$20 range in line strengths covered by the Paschen series.}
    \label{p1:fig:B331_individual_line_fits}
\end{figure*}

\subsubsection{Hydrogen emission lines}

Most of the hydrogen emission line profiles are clearly double peaked and show relatively broad line wings. The latter correspond to high rotational velocities of up to $\sim 200$ km\,s$^{-1}$. This suggests a Keplerian rotating disk extending down to the stellar surface. 
We find a best fitting inner radius of <0.12\,AU (<1.2\,$R_\star$) for nearly all lines. The scale height at the inner rim is quite similar as for B243 and points to the inner disk being puffed up.
The surface density at the inner radius is well constrained at $1.1 \times 10^{3}$\,g\,cm$^{-2}$. A similar column density has been found for NGC 2021 IRS  based on CO emission \citep{2020A&A...635L..12G}. 

For B331 too the observed Paschen and Bracket series span a large range in line strengths, of about a factor of 20 (Fig. \ref{p1:fig:B331_individual_line_fits}). The dashed blue lines in Fig.~\ref{p1:fig:B243_B331_parameter_trend} show the best fit parameters taking into account the same set of Paschen lines as for B243 and, in addition, Br-11 to Br-16. With this set, the line strengths are again reproduced to within a factor of two save for Pa$\beta$ and Br$\gamma$ for which the difference is somewhat larger; see the orange line in Fig.~\ref{p1:fig:B331_individual_line_fits}. None of the models within the confidence interval reach the upper limits.
The best fit yields an inclination of about $75^{\circ}$.

Similar to B243 we find a systematic trend in that the observed line strengths vary less with Paschen and Bracket line number than do the model predictions. We return to this in Section~\ref{p1:section:discussion}. 

\subsubsection{SED}
The SED of B331 shows an IR excess at $\lambda > 3\,\mu$m; see Fig.~\ref{p1:fig:SED_fit}. We use the best fit inclination of the inner disk of 75$^\circ$. The \prodimo\ dust model results in a maximum temperature of $\sim$400~K positioning this warm dust at $\sim$100~AU from the star. 
The outer radius of the dusty disk remains unconstrained, simply because we lack photometric or spectroscopic constraints at wavelengths well beyond 10\,$\mu$m. 
The amount of 3-10\,$\mu$m flux is a function of the scale height of the inner rim of the outer dusty disk. We find for this scale height 0.18$^{+0.02}_{-0.02}$\,AU.  The surface density at this location should be at least 19\,g\,cm$^{-2}$, with improving fits for higher values up to at least $\sim$700\,g\,cm$^{-2}$. 
\\

In summary, the overall disk properties of B331 are quite different from those of B243. B331 lacks hot dust; warm dust seems only present starting at about 100\,AU.  This suggests that for this source a disk gap is created. The inner boundary of this gap, if present, is poorly constrained (the hydrogen spectral lines indicate it should start at a distance $> 2$\,AU). The properties of the inner gaseous disk up to 3.5\,AU appear quite similar to B243, both having a puffed-up inner rim.

\subsection{2D Disk structure of B243 and B331} \label{p1:subsection:physical_properties}

\begin{figure*}
    \centering
    \includegraphics[width=0.7\textwidth]{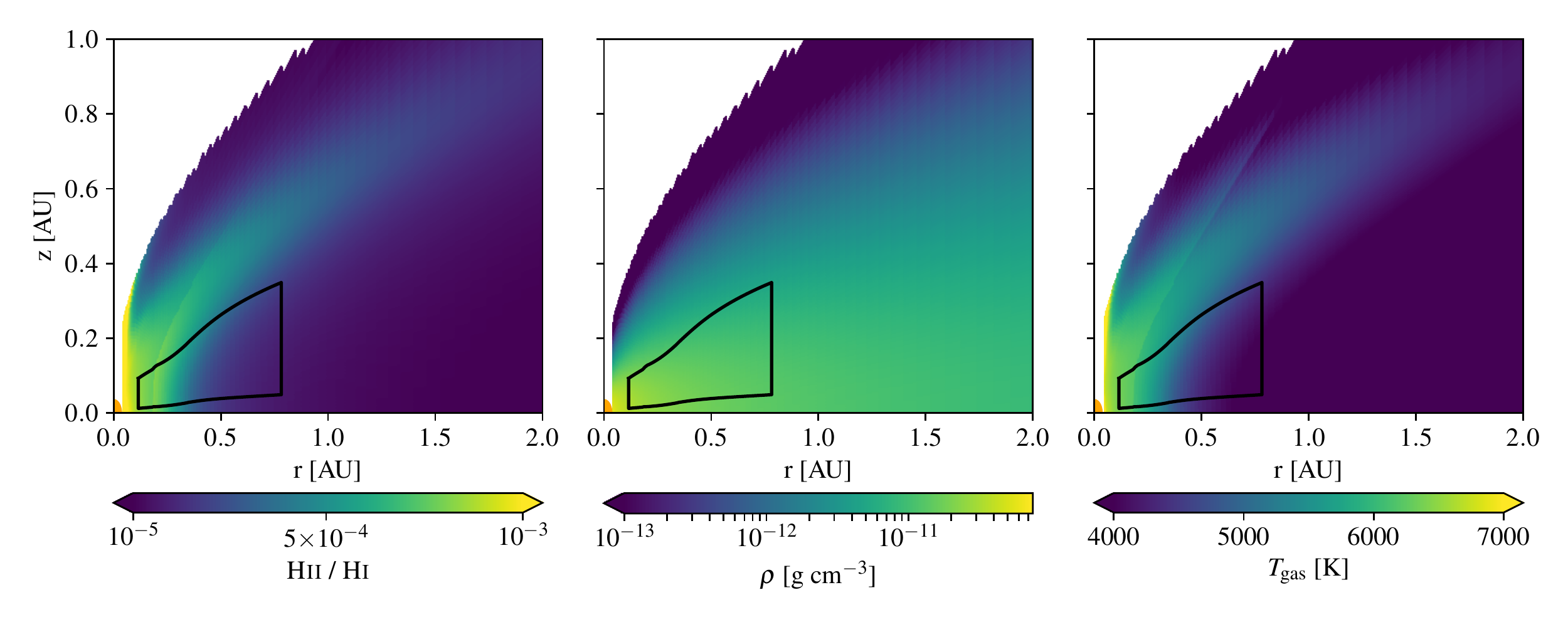}
    \caption{Hydrogen ionization, density, and temperature structure of the inner region of B243. The parameters used match the values indicated by the full red line in Fig.~\ref{p1:fig:B243_B331_parameter_trend}. The black contour indicates the approximate origin of 50\% of the Pa-16 line emission based on vertical escape probabilities. The orange circle on the left indicates the central object. }
    \label{p1:fig:B243_physical_properties}
\end{figure*}

\begin{figure*}
    \centering
    \includegraphics[width=0.7\textwidth]{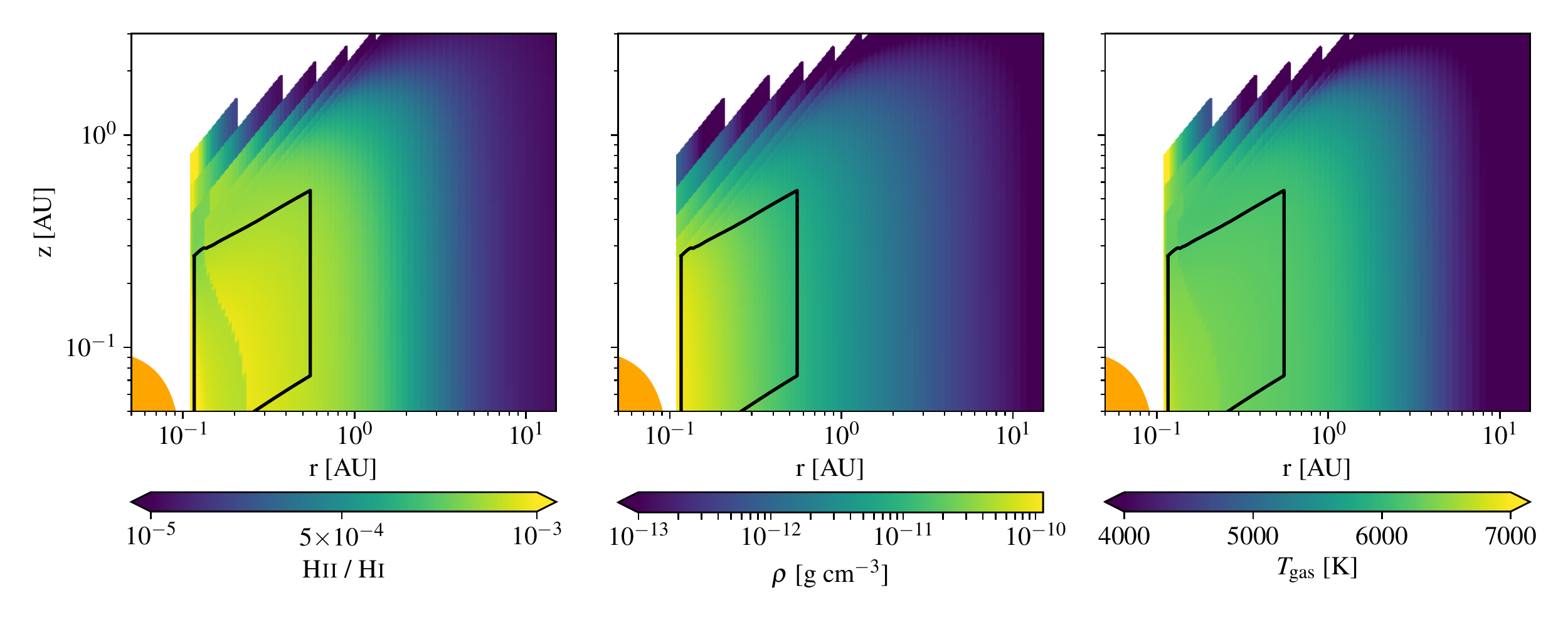}
    \caption{Same as Fig.~\ref{p1:fig:B243_physical_properties}, but now for B331 on a logarithmic spacial scale.
     The parameters used match the values indicated by the dashed red line in Fig.~\ref{p1:fig:B243_B331_parameter_trend}. 
     }
    \label{p1:fig:B331_physical_properties}
\end{figure*}

Figures \ref{p1:fig:B243_physical_properties} and \ref{p1:fig:B331_physical_properties} show the hydrogen ionization fraction, density, and temperature  structure of the best fitting inner disk models for B243 and B331, respectively.
The distance axis is linear in the case of B243 and logarithmic in case of B331.
The hydrogen ionization fraction throughout the disks is low, even at the inner rim of the gaseous disk (where it reaches $\la 10^{-3}$). This is essentially due to the fairly low effective temperatures of the two stars. The ionizing photons do not penetrate far into the disk as the mostly neutral (hydrogen) medium becomes quickly optically thick for Lyman continuum radiation. B331 is almost an order of magnitude more luminous than B243, which is why for this star the thin somewhat ionized shell reaches deeper disk layers. For the same reason, the disk of B331 is heated to higher temperatures further out.

The black contours in the figures indicate the origin of half of the Pa-16 line flux. The formation region stretches out to about a radial distance somewhat less than one AU, that is hydrogen lines probe the very inner part of the disk only -- dominated by the puffed up inner zone. Similar to the work by \citet{2021A&A...654A.109K}, \citet{2016A&A...589L...4C}, and \citet{2010Natur.466..339K}, we find the hydrogen emission to originate from a region closer to the central star than the continuum emission, based on the location of the inner radius of the dust disk. We find the extent of the hydrogen emission region to reach up to $\sim$1~AU, which is in line with the most compact emitting regions found in interferometric studies of \citet{2021A&A...654A.109K} and \citet{2016A&A...589L...4C}, but is more compact than what they typically derive.

\section{Discussion}
\label{p1:section:discussion}

\begin{figure}
    \centering
    \includegraphics[width=\columnwidth]{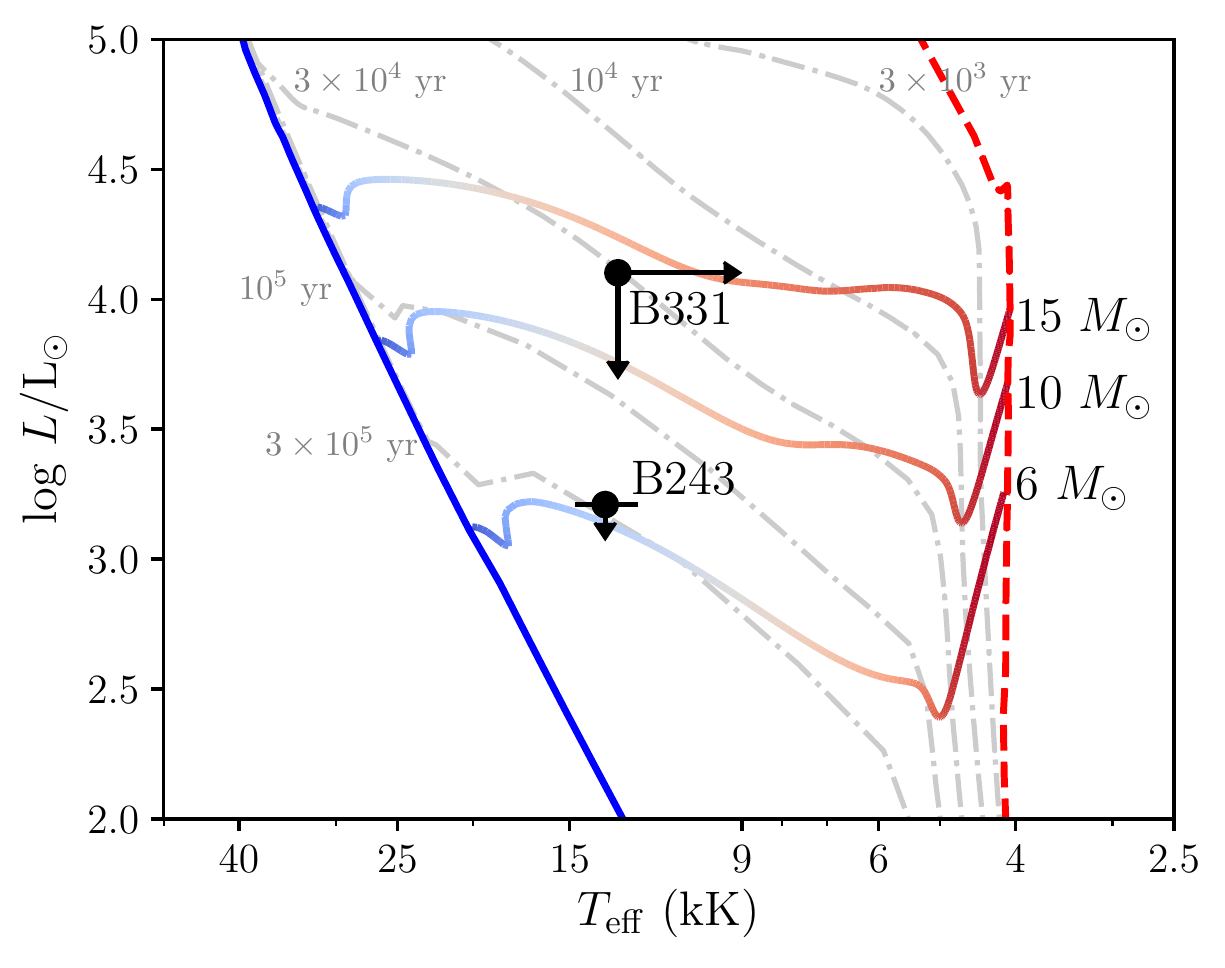}
    \caption{Adapted from Fig.~9 of \citepalias{2017A&A...604A..78R}. Positions of B243 and B331 in the Hertzsprung-Russel Diagram. The solid blue line on the left indicates the zero age main sequence. The dashed red line on the right indicates the birth line. The solid line connecting the birth line and the zero age main sequence indicate the MIST pre-main sequence evolutionary tracks of stars with masses indicated on the right \citep{2016ApJS..222....8D}. The colour of this line indicates the relative time between birth and reaching the main sequence. The dash dotted grey lines indicate the isochrones corresponding to $3\times10^3, 10^4, 3\times10^4, 10^5$ and $3\times 10^5$ years from top to bottom respectively. }
    \label{p1:fig:HRD}
\end{figure}

Before summarizing and discussing the disk properties of the M17 members B243 and B331, we show their positions in the Hertzsprung-Russell diagram (HRD) in Fig.~\ref{p1:fig:HRD}. \citetalias{2017A&A...604A..78R} estimate an age for the M17 star-forming region of less than 1\,Myr. This, combined with both sources showing both stellar absorption and disk emission features, strongly suggests that the stars are pre-main-sequence objects on Thomson or Henyey (a.k.a. radiative) tracks; in the final phase of their formation contracting towards the zero-age main sequence. The luminosity (and associated mass of $\sim$12\,$M_{\odot}$) classifies B331 as an MYSO source. The estimated mass (of $\sim$ 6\,M$_{\odot}$) of the lower luminosity source B243 is more representative of the high-mass end of Herbig Be sources. The total MYSO lifetime, defined as the time from passing the birthline until arrival on the main sequence, is less than $10^{5}$\,yr for B331 and a few times $10^{5}$ yr for B243 \citep{2009ApJ...691..823H}; both stars have progressed considerably in this evolution. Given these short evolutionary time scales and the observation that higher mass main-sequence stars in M17 lack disk signatures \citepalias{2017A&A...604A..78R}, one may anticipate that the disks of B331 and B243 are in the process of being cleared.

\begin{figure}
    \centering
    \includegraphics[width=\columnwidth]{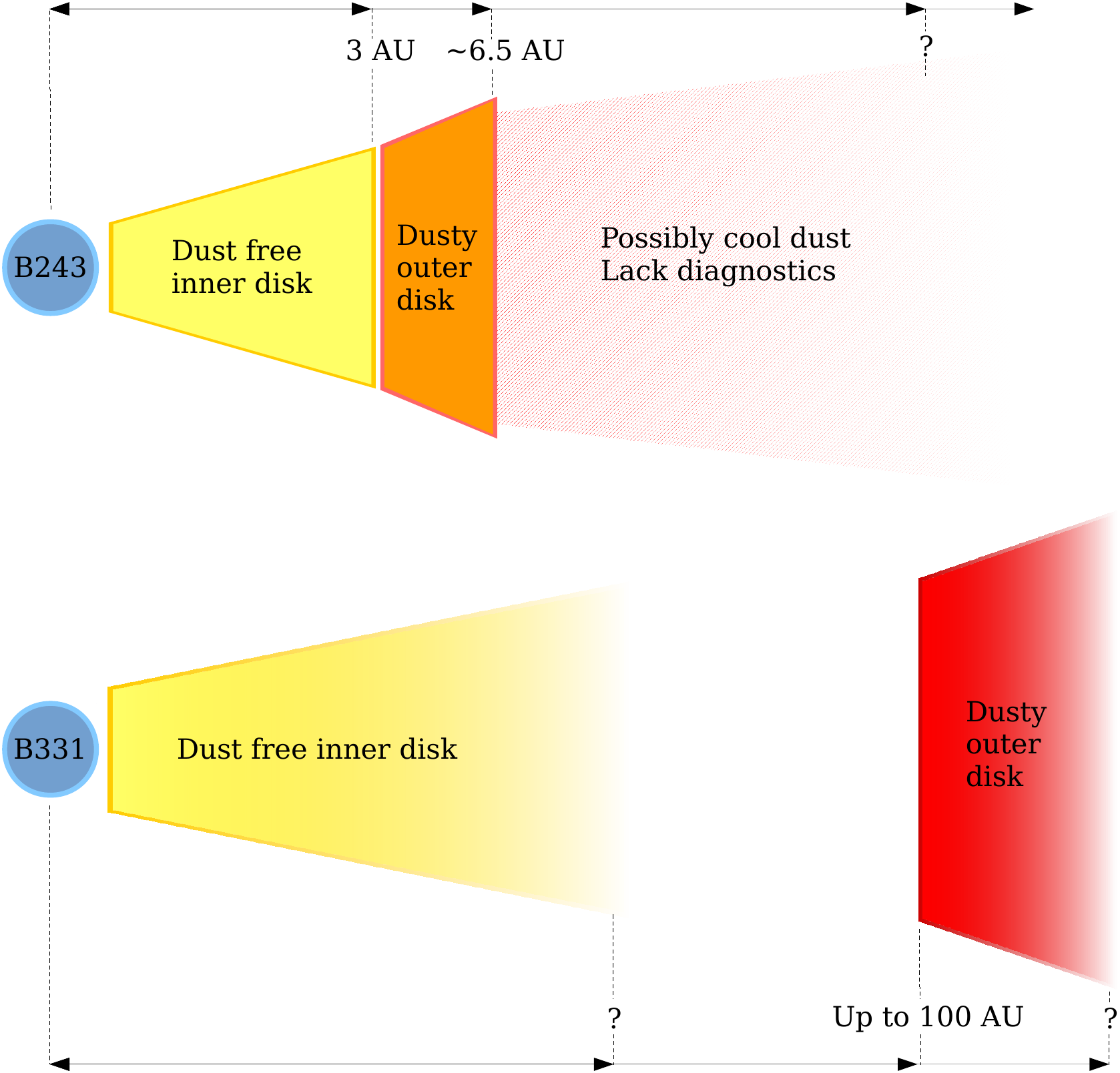}
    \caption{Schematic overview of the geometric properties of the disks of B243 and B331.}
    \label{p1:fig:schematic_overview}
\end{figure}

A schematic diagram illustrating the main properties derived for the two disks is shown in Fig.~\ref{p1:fig:schematic_overview}. Both stars feature a gaseous disk that almost (within a stellar radius or $\sim$0.1\,AU) reaches the stellar surface. The hydrogen emission lines originate from within the first AU of these disks, signifying that the full extend and properties of the gaseous disk are not probed by these diagnostics.
The disk of B243 contains hot dust of $\sim$1500~K, whereas the hottest grains surrounding B331 have a much lower temperature of 400\,K implying a significant dust free inner zone spanning $\sim$100~AU. The extend of the dust free inner zone depends on the continuum optical depth of the gaseous disk. A very opaque gas disk would move the dust disk closer to the star. The best fit model for the dust disk of B243 suggests an outer radius of $\sim$6.5~AU, however, this is poorly constrained.
The SED modelling suggests B243 is consistent with having a Group\,{\sc II} disk and B331 a Group\,{\sc I} disk.
However, as our diagnostics are not sensitive to cool dust (our longest wavelength point is at 8\,$\mu$m), we cannot rule out the presence of a (large) gap and cool dusty outer disk or a self-shadowed region in B243. 
Without the 8\,$\mu$m point the outer radius of the hot dust disk would not be constrained. We conclude that both disks appear disrupted and identify them as transitional disks. Though not subject to analysis, both disks show pronounced CO-bandhead emission (\citetalias{2017A&A...604A..78R}, Poorta et al. in prep.).
 
The gaseous inner disks of both stars have significantly larger scale heights close to the star than would be expected from standard hydrostatic calculations, consistent with puffed-up inner rims \citep{2004A&A...417..159D}. This is further discussed in Section~\ref{p1:sec:scale_height_discussion}. 
We adopt a Keplerian rotation profile and did not find any indication that parts of the disk regions probed by the Paschen lines experience significant sub- or super-Keplerian motion. Still, we cannot fully exclude deviations from Keplerian motion as the rotation velocity is somewhat degenerate with the inner radius and inclination of the disk.

The small gaseous disk inner radii (of less than 2\,$R_\star$) differs from the situation in lower mass YSOs. In these lower mass counterparts the disk material does not directly reach the stellar surface, but gas loops towards the surface via magnetoshperic accretion \citep[e.g. ][]{2013ApJ...767..112I}. This leaves a gap between the disk and star. \citet{2015MNRAS.453..976F} study a large sample of Herbig AeBe stars and find the UV excesses of early type Be stars to be inconsistent with magnetospheric accretion \citep[see also][]{2007MNRAS.376.1145W}. This suggests that higher mass stars lack strong magnetic fields. 
Another indication for the absence of a strong magnetic field is consistent with Keplerian rotation. Strong magnetic fields can cause the disk to co-rotate with the stellar surface. This results in lower than Keplerian velocities close to the star and would likely result in too low velocities to reproduce the observed line profiles. 
In conclusion, the disks reaching almost the stellar surface and being compatible with Keplerian rotation close to the inner rim is in line with a lack of magnetospheric accretion, consistent with the general consensus that most higher mass stars lack a strongly magnetized envelope \citep{2007MNRAS.377.1363M,2017MmSAI..88..605O}.

\subsection{Uncertainties in stellar properties}
Table \ref{p1:tab:Stellar_parameters} shows the stellar parameters of B243 and B331. For B243 they result from quantitative spectroscopy; for B331 from SED fitting. The main uncertainties on the properties of these stars are the temperature, extinction ($A_V$ and $R_V$), and luminosity. The extinction parameters do not affect this work significantly as we fitted the normalized line profiles, and the extinction at the NIR is modest. The luminosity affects the dust sublimation radius of B243 and location of warm dust for B331. A higher luminosity also results in higher hydrogen line flux (but not necessarily a higher \emph{normalized} line flux, as the continuum flux is also increased). 

The surface gravity of B331 is unconstrained; we assume a value $\log g = 4.0$. The gravity of the star mostly affects the wings of the circumstellar H\,{\sc i} emission profiles through the normalization process (see section\,\ref{p1:subsection:norm_and_corr}), a too high gravity giving a broader emission profile. This would have to be compensated by a lower scale height or smaller inner radius. 

\subsection{Puffed up inner rim}\label{p1:sec:scale_height_discussion}
At the inner rim, both B243 and B331 have a scale height significantly larger than expected based on the hydrostatic equilibrium approximation $H \sim c_{\rm s} / \Omega_{\rm K}$, with $c_{\rm s}$ the sound speed and $\Omega_{\rm K}$ the Keplerian rotation frequency. For $T=6000$\,K, $r=0.2$\,AU and $M=6$\,M$_\odot$ the adopted scale height for B243 is 15 times the hydrostatic value. This 'puffed' up region is unlikely to extend far in to the disk. Despite such puffed up regions have been found before in efforts to model the inner region of disks \citep[e.g.][]{Woitke09,2001ApJ...560..957D}. A possible explanation for the large scale height could be that a disk wind contributes to the observed emission. However, a disk wind would possibly have a velocity profile distinct from a puffed up Keplerian disk. However, the velocity may still be dominated by orbital motion if the emission originates from close to the disk, with only a slight broadening of the profile.

We also investigated the effect of mass accretion on the inner rim structure. To this end, we performed test calculations of the best fit models to both B243 and B331 in which we use an accretion rate of 10$^{-4}$\,M$_\odot$\,yr$^{-1}$. In these cases the line emission gets significantly stronger with 5 and 1.5 times more line flux for B243 and B331, respectively. The increase in line strength could be compensated by lowering the scale height. This would require a significant gas reservoir to be present to feed these accretion flows.

\subsection{Origins of the hydrogen line emission and the role of chemistry}

The hottest parts of the disks of B243 and B331 reach temperatures of $\sim$7000\,K and hydrogen ionization fractions of $\sim 10^{-3}$.
We find charge exchange reactions to be the main source of hydrogen ionization. These reactions are facilitated by various heavier elements, in particular sulfur. Therefore, the abundance of these metallic species are of importance to the line formation. A lower metallicity will result in weaker lines. 
We note that the reaction rate coefficient of the charge exchange reaction with sulfur is highly uncertain. In \prodimo\ an estimate for the reaction rate coefficient of $5 \times 10^{-12}$\,cm$^3$\,s$^{-1}$ is supplied. However, \citet{1980A&A....85..144B} find an upper limit of $3\times 10^{-15}$\,cm$^3$\,s$^{-1}$ at a temperature of $10^4$\,K for the reaction rate coefficient. The hydrogen emission line strength scales nearly linearly with the reaction rate coefficient, with a reaction rate reduced by a factor of 50 resulting in 40 times weaker lines. This would significantly affect the derived properties of the disk, requiring substantially higher disk masses. 

Presently, the central stars' temperatures are too low to produce sufficient ionizing photons. Therefore, we conclude that the stars appear to be in a phase of formation in which chemistry dominates the production of ionized hydrogen. We note that as the stars move closer to the zero-age main sequence (ZAMS) photo-ionization may take over as the main mechanism of hydrogen ionization, particularly so for B331 (which may reach a ZAMS temperature of about 30\,000\,K), if a close-in gaseous disk should prevail to such a late stage.

The key role of chemistry also highlights the importance of the chemical network and the corresponding reaction rates. In this work we use the {\sc UMIST\ 2006} reaction network. We note that employing the {\sc UMIST\ 2012} reaction network results in $\sim$20\% weaker hydrogen emission lines. The charge exchange reaction rate coefficients of hydrogen and sulfur are not included in these networks, but instead are supplied by a separate \prodimo\ input. We also do not find a significant difference in hydrogen line emission between the fiducial large and small chemical networks available in \prodimo\ \citep{Kamp17}.

\subsection{Origin of the hydrogen line emission and accretion}

Hydrogen emission lines are generally considered to be tracers of accretion when observed in (massive) young stellar objects \citep[e.g.][]{2014MNRAS.445.3723I, 2017MNRAS.464.4721F, 2011A&A...535A..99M}. This raises questions as to the possibility and nature of accretion in the disks. \citet{2017MNRAS.464.4721F} speak of a correlation of line luminosities with accretion rate, while also noting that accretion need not be the physical origin of the lines. Their empirical relation between hydrogen line strength and accretion rate implies an accretion rate of $\sim$10$^{-5}$ and $\sim$10$^{-3}$\,M$_\odot$\,yr$^{-1}$ for B243 and B331, respectively. If a gap of at least tens of AU indeed separates the inner disks from more distant disk material (if present at all), these rates would deplete the inner disks, up to a few AU, of the order of years. The very small chance of detecting the systems at the exact moment of inner disk accretion suggests that either the inner disk is being fed from a more extensive outer reservoir, or, that the hydrogen emission is not an accurate tracer of accretion in these systems.

To expand on the latter possibility, the empirical relation mentioned above is derived using calibrations based on magnetospheric accretion modelling. Following from our previous discussion of disk geometry and magnetic fields, as well as drawing from recent literature on accretion luminosities in Herbig AeBe stars \citep[e.g.][]{2020MNRAS.493..234W, 2022ApJ...926..229G,2020Galax...8...39M} it is likely that higher mass YSOs ($\gtrsim$4\,M$_{\odot}$) accrete by a different mechanism. An obvious alternative is boundary-layer accretion, where disk material slows down in a transition layer between the disk and the star, releasing its kinetic energy in that boundary layer. \citet{1974MNRAS.168..603L} relate the mass accretion rate to a black body emission from an annulus with temperature $T_{\rm BL}$. They assume all orbital kinetic energy is dissipated through radiation from the boundary layer. However, this assumes the layer to be optically thick and the radiation from the central star heating up the material is ignored. Despite those assumptions, using the accretion rates above we find $T_{\rm BL}$ to be larger than the temperature in the inner rim of the disk for both B243 and B331. A more detailed modelling of boundary-layer accretion including the stellar radiation field and more accurate heating and cooling processes would be required to accurately link the temperature of the inner disk and the accretion rate of the star. Currently, there is no model that relates (hydrogen) line emission to accretion rate in the BL regime.

\subsection{Disk effect on the central star}
The disks, given their small inner radii and the possibility of ongoing accretion, might still affect the properties of their central star on its way to the main-sequence. 

Following the arguments made in the previous section and based on the derived disk masses, it is likely that the central stars have accumulated the mass at which they will start central hydrogen fusion.
However, the inner disk mass may contain a sizable amount of angular momentum. Here we estimate the effect of the disk on the surface rotational velocity of the star once it reaches the ZAMS.

Assuming conservation of angular momentum for the stars and no further accretion we can calculate the surface rotational velocity upon arrival on the ZAMS. On the basis of their assumed stellar masses, 
we adopt radii of 2.5~R$_{\odot}$ and 4.2~R$_{\odot}$ at the ZAMS for B243 and B331, respectively \citep{2011A&A...530A.115B}. The radial density structure of the stars is approximated using the solution of the Lane-Emden equation for a polytropic index of 1.5 in both the current state and at the ZAMS. We further assume solid body rotation. Taking an inclination of $75^\circ$ converts the $v\sin i$ of 110~km\,s$^{-1}$ into $v_{\rm rot} = 114$~km\,s$^{-1}$ for the present-day state of B243. After contracting to the main sequence this results in a rotational velocity of 342~km\,s$^{-1}$, which is $\sim$0.50\,$v_{\rm crit}$, where $v_{\rm crit}$ is the critical rotation rate. If we adopt a column density at the inner radius of the disk of 1000~g\,cm$^{-2}$ and a relatively flat column density exponent of $\epsilon = -0.5$, the angular momentum of a 5\,AU inner disk would be $\sim$17\% of that of the star. The current $v_{\rm rot}$ has not been constrained for B331. Doing the same calculation for this star under the same assumptions and taking the current $v\sin i = 110$~km\,s$^{-1}$, the same value as for B243, yields a ZAMS spin velocity of $\sim 592$~km\,s$^{-1}$, which is $\sim$0.80\,$v_{\rm crit}$. The angular momentum of the disk is then $\sim$~7\% of that of the star.

These estimates suggest that the inner disk is unlikely to still contribute strongly to the ZAMS rotational velocity of the central star; its value is essentially controlled by stellar contraction. Spin velocities of 0.5-0.9 $v_{\rm crit}$ upon arrival on the main sequence are in line with previous findings for stars in the same mass range \citep{2010ApJ...722..605H}. For stars with a weak or lacking magnetic connection to the disk (see above) gravitational torques are expected to limit the ZAMS spin velocity to about half critical \citep{2011MNRAS.416..580L}. This may be an indication that the current projected spin rate of B331 is less than the adopted 110\,km\,s$^{-1}$.

\subsection{Disk disruption scenarios}
The leading mechanisms responsible for the dispersal of disks around young stars are photo-evaporation \citep[e.g.][]{2009ApJ...690.1539G,2010MNRAS.401.1415O}, stellar or planetary companion formation \citep[e.g.][]{2013A&A...560A..40M}, and possibly stellar winds for the most massive YSOs \citep{2006A&A...455..561B}. 
Given the relatively high temperatures of the MYSO sources studied here relative to pre-main sequence stars of lower mass, photo-evaporation may be a contender for disk dispersal. \citet{2010MNRAS.401.1415O} provide a simple scaling relation for the characteristic radius where thermal evaporation by EUV light starts, after which more inner parts of the disks are cleared on a timescale of a few times $10^{4}$~yr assuming the disk dispersal time is only a weak function of stellar mass as suggested by \citeauthor{2010MNRAS.401.1415O}. For our sources this radius is at about 100\,AU; it corresponds well to the size of the dust free gap or zone we observe in both sources. In this mechanism, the innermost parts of the disks (i.e. those probed by H lines) survive the longest \citep{2006MNRAS.369..229A}.
So, photo-evaporation as a disk dispersal mechanisms seems congruent with the derived disk properties for both B243 and B331. We remark however that photo-evaporation models do not take into account a puffed-up inner disk, as found for both stars, that may extinct a sizeable amount of the ionizing radiation, limiting the EUV-illumination of further out regions and hence efficient disk dispersal. \citet{2009ApJ...690.1539G} study the effect of X-ray, EUV, and FUV radiation on the photo-evaporation of disks. They find the photospheric FUV radiation to be the dominant energy source driving mass-loss from the outer regions of the disk, and EUV radiation to only affect the inner regions of the disk. We expect the inner gaseous disk to be optically thin to FUV radiation and the central stars to be bright in the FUV, therefore it would be a potential energy source for photo-evaporation. 

Ongoing photo-evaporation would result in observable spectral features such as forbidden oxygen emission and H$_2$ emission \citep[e.g.][]{2020A&A...643A..32G}. Though we have not detected any H$_2$ emission, B243 shows forbidden [O{\sc i}] $\lambda 630$~nm emission. Its behaviour is different on-source than off-source pointing at a possible circumstellar origin (Derkink et al. in prep.). The [O{\sc i}] emission of B331 does not allow us to make this distinction.

Gaps in proto-planetary disks are commonly observed in the low mass counterparts of MYSOs and were first identified in \citet{1989AJ.....97.1451S}. \citet{2013ApJ...769..149K} collate 105 of these 
disks. They propose the gaps are likely due to companions orbiting the central object, and that an origin linked to grain growth, turbulence or photo-evaporation processes is less likely. 

Companions formation through gravitational instabilities in efficiently cooling disks appears feasible on the timescales of MYSO formation, at least theoretically \citep[e.g.][]{2003MNRAS.346L..36R,2020A&A...644A..41O}. In this scenario the companion will create a gap in the disk, preventing material to flow through its orbit efficiently.
A small amount of gas may travel past the gap to sustain the observed gaseous inner disk while blocking dust, creating a dust free inner disk \citep[e.g.][]{2006ApJ...641..526L}. This scenario too seems in line with the derived disk properties for both stars. We note that such companion forming disks are expected to develop a spiral structure, which, if indeed the correct scenario, may perhaps play a role in explaining the (modest) discrepancies between predicted and observed H line profiles discussed below.

We conclude that both disk dispersal mechanisms appear reasonable candidates to explain the observed disk structures in B243 and B331.

\subsection{Deviating trends in hydrogen series properties}

Figures~\ref{p1:fig:B243_individual_line_fits} and~\ref{p1:fig:B331_individual_line_fits} show that a single model cannot fully recover the line strengths of all modelled Paschen and Brackett series lines: lower (higher) series lines are predicted stronger (weaker) than observed. Differences remain within a factor of two though, while the range of line strengths spans a factor $\sim$20$-$30. In addition to this, we observe a modest discrepancy in the trend of peak separation versus line strength. 

\begin{figure}
    \centering
    \includegraphics[width=\columnwidth]{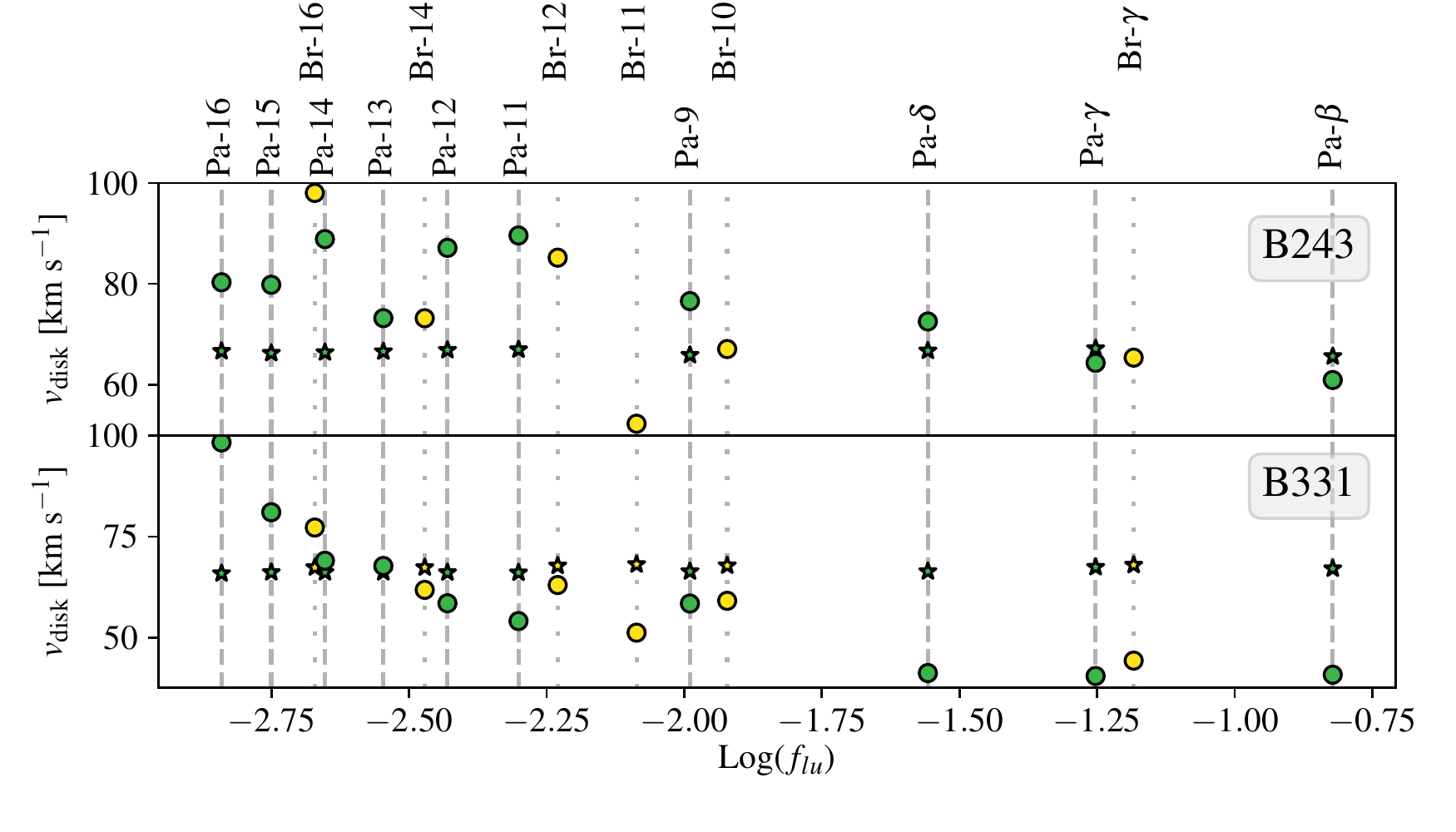}
    \caption{Orbital velocity of the hydrogen gas based on the peak separation of the observed (circles) and simulated (stars) emission lines as function of their oscillator strength. Green markers indicate Paschen lines and yellow markers Brackett lines. The velocity is determined by fitting double gaussians to the profile. The simulated profiles are from Figures~\ref{p1:fig:B243_individual_line_fits} and \ref{p1:fig:B331_individual_line_fits}. }
    \label{p1:fig:disk_velocity}
\end{figure}

Figure~\ref{p1:fig:disk_velocity} shows that the disk velocity decreases for increasing oscillator strength for both B243 and B331. As in our Keplerian disk model all line emission originates from the optically thin inner disk region (i.e. from the first AU) no such trend is predicted.
For classical Be stars a similar trends in found as in our systems \citep[e.g.][]{2012ApJ...744...19K}. These authors could spatially resolve the emitting region of different groups of hydrogen lines for one such Be disk and find spatially different formation regions. This indicates that in this Be disk the hydrogen lines are optically thick (unlike in our MYSO modelling).

Though we cannot pinpoint the cause of the mismatch in peak separation trend, part of it may possibly be explained by deviations from azimuthal symmetry in the inner disk. It also remains to be investigated how the inclusion of (boundary-layer) accretion and disk winds would affect the modelled lines. Massive YSO sources in M17, including the objects studied here, show variability in their spectra originating in their circumstellar disks (Derkink et al. in prep.). This variability includes excess line emission moving from the red to the blue part of the line on timescales consistent with the rotation period of the inner disk, underlining that likely there are processes active in the inner disks that break axial symmetry. Possible features include spirals, warps, large scale clumps or other, more chaotic, forms of asymmetries. 

\section{Summary}\label{p1:section:summary}

We have investigated the inner disk regions of two MYSO sources, B243 (with an estimated mass of 6\,\Msun) and B331 (12\,\Msun) in the star-forming region M17, using the thermo-chemical code \prodimo. Likely, these inner disks are remnant structures from the star assembly process.

Our main diagnostics are (double-peaked) hydrogen lines of the Paschen and Brackett series and near-IR photometry. The first allow us to probe the kinematics and structure of the gaseous inner disk and the second the thermal emission of hot dust. The effective temperatures of the central stars are such that photo-ionization of hydrogen is much less important than ionization through charge exchange reactions, so it is chemistry that dominates the H line-formation.

Our main findings are:
\begin{enumerate}
\item A small puffed-up gaseous disk extends to very close to the stellar surface in both Group\,{\sc II} source B243 and Group\,{\sc I} source B331. In B243 a dust free cavity of 3~AU is present with hot dust of 1500~K at the inner rim of the dusty disk. We find a best fit outer radius of $\sim$6.5~AU, but this is poorly constrained. The inner disk of B331 is dust free and probably of similar dimension, cool dust indicating an outer disk starting at about 100\,AU with a dust free (and possibly gas free) gap in between. 

\item The inner disk extending to almost the stellar surface suggests that some accretion might still be ongoing, likely through boundary-layer (BL) accretion. Magnetospheric accretion is less likely as the geometry of this accretion mechanisms suggests an inner gaseous disk gap. A lack of magnetospheric accretion is in line with the general consensus that a different accretion mechanism is at work in higher mass stars. A first order approach indicates that it would be interesting to investigate the BL mechanism for higher mass Herbig Be stars and MYSOs with detailed line modelling. The presence of disk winds remains an open question.

\item The inner disk contains too little mass and too modest angular momentum to significantly change the final (i.e. ZAMS) mass and final spin velocity of the stars. Concerning the latter, contraction towards the main sequence is the main effect impacting the spin velocity in the remainder of pre-main sequence evolution \citep[see also][]{2017A&A...604A..78R}. The angular momentum of the inner disk is of the order of 10\% of that of the central star. 

\item The disk structures of both sources strongly suggest that the disks are in the process of being cleared, that is they are transitional disks. Possible disk dispersal mechanisms are photo-evaporation and stellar or planetary companions formation. The derived properties of the disks are compatible with both scenarios, specifically the presence of a small inner disk and a disk gap towards a more distant ($\sim$100 AU) outer disk (in B331 and possibly in B243).

\end{enumerate}

The study presented here does not fully characterize the disks orbiting B243 and B331; but it does present the first detailed 2D thermo-chemical radiative transfer modelling of hydrogen lines in such sources. High resolution, longer wavelength imaging with for example ALMA would greatly add to our insight in the properties of the outer regions of the disk. Whether such regions (if they exist at all in B243) would still play a role in the formation process of the central object remains to be seen, given the proximity of the stars to the ZAMS. The fate of such outer disks may simply be that they are dispersed by the concerted action of the star's H\,{\sc ii} region and stellar wind upon arrival on the ZAMS \citep[e.g.][]{2021MNRAS.501.1352G}. Studies of line variability may contribute to our understanding of dynamical process in the inner disks, possibly due to the presence of companions. The combination of these different approaches may greatly help in unravelling the architecture of companion systems around massive stars, their possible migration \citep{2021A&A...645L..10R}, and pre-main sequence or early main sequence merging with the primary star \citep{2022NatAs...6..480W}; topics that constitute new and exciting problems in massive star formation.

\section*{Acknowledgements}\label{ack}

We thank the anonymous referee for insightful comments that helped to improve this manuscript. This publication is part of the project ‘Massive stars in low-metallicity environments: the progenitors of massive black holes’ with project number OND1362707 of the research TOP-programme, which is (partly) financed by the Dutch Research Council (NWO). This work is based on observations collected at the European Organisation for Astronomical Research in the Southern Hemisphere under ESO programme 089.C-0874(A). We thank SURF (www.surf.nl) for the support in using the Lisa Compute Cluster.

This research has made use of NASA’s Astrophysics Data System.  This work makes use of the Python programming language\footnote{Python Software Foundation; \url{https://www.python.org/}}, in particular packages including NumPy \citep{harris2020array}, SciPy \citep{2020SciPy-NMeth}, and Matplotlib \citep{Hunter:2007}.

\bibliography{bibfile}

\appendix

\section{Normalized and corrected spectra}
\begin{figure*}
    \centering
    \includegraphics[width=\textwidth]{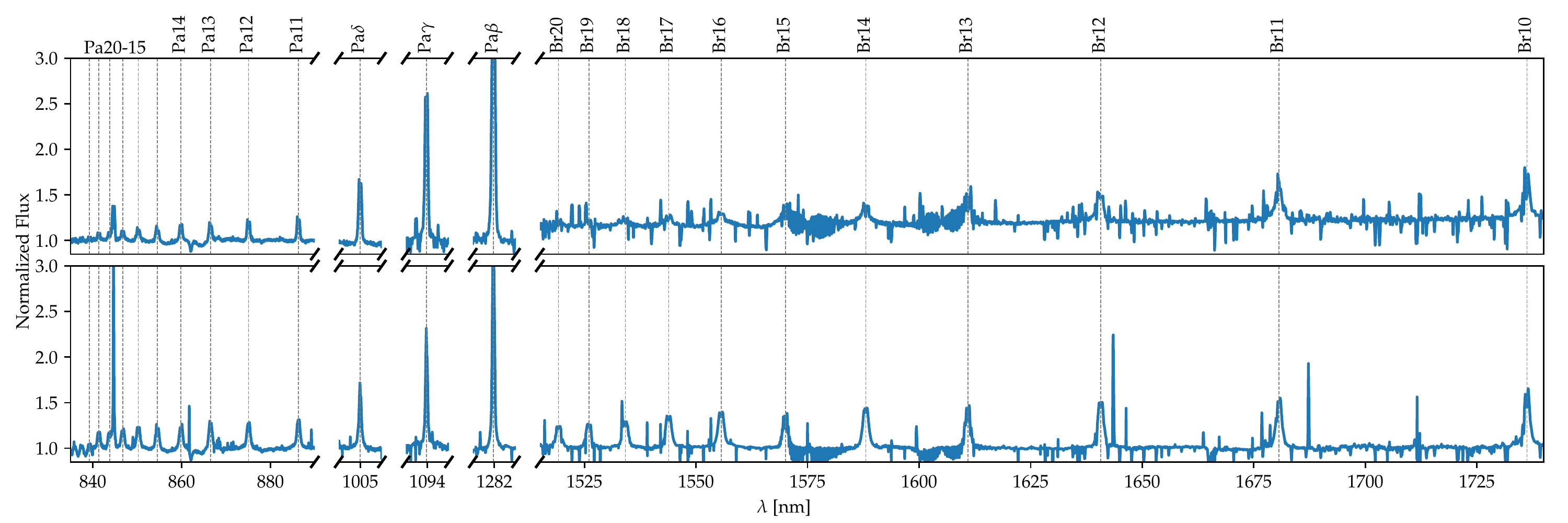}
    \caption{Normalized and corrected spectra of B243 (top panel) and B331 (bottom panel), focusing on the Paschen and Brackett series. The spectrum of B243 is not normalized at wavelengths longer than 1500\,nm due to the normalization method's incompatibility with the IR excess. }
    \label{p1:fig:almost_full_spectra}
\end{figure*}

The normalized and corrected spectra of B243 and B331 are shown in Fig.~\ref{p1:fig:almost_full_spectra}. The spectra have only been normalized around the modelled lines. The continuum of B243 at wavelengths longer than 1500\,nm deviates due to IR excess from dust emission.

\section{Model test -- HII region}
\begin{figure}[h!]
    \centering
    \includegraphics[width=\columnwidth]{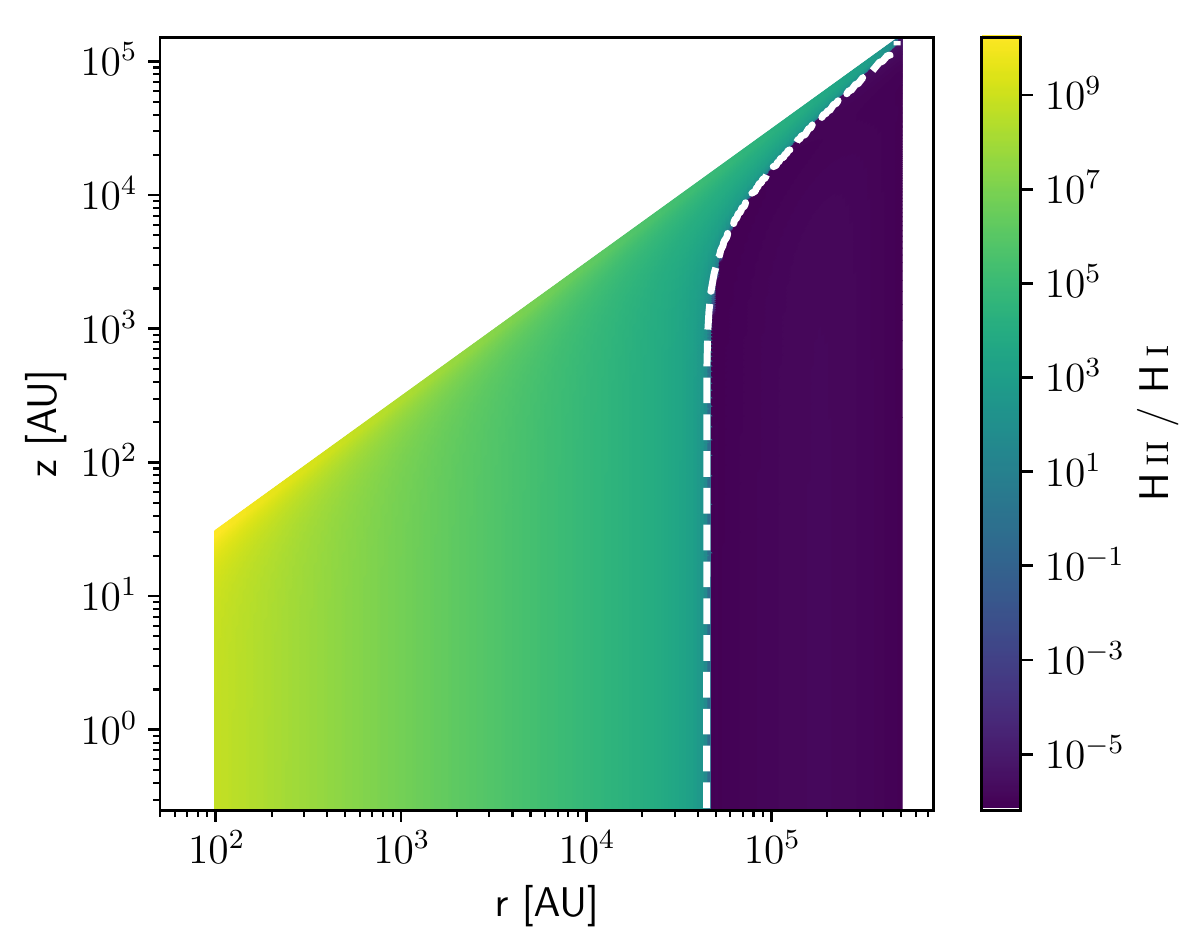}
    \caption{H\,{\sc ii} region simulated with \prodimo. The colour indicates the ratio between neutral and ionized hydrogen. The white dashed line indicates the analytically calculated Str\"omgren radius based on the radiation field, temperature and density of the medium.}
    \label{p1:fig:HII_contour}
\end{figure}

The new hydrogen atom description and extended radiation field were tested by simulating an H\,{\sc ii} region and comparing the numerically calculated ionization edge with the analytical Str\"omgren radius. We simulate an O-type star with $T=38\,000\,$K and $L_\star=1.65\times 10^5$\,L$_{\odot}$. We surround this star with large gaseous disk with a constant radial density in the mid-plane ($\rho \sim 10^{-20}$\,g\,cm$^{-3}$). The disk does not contain dust or poly-cyclic aromatic hydrocarbons (PAHs). The disk material has (proto-)solar abundances. Fig.~\ref{p1:fig:HII_contour} shows the ionization fraction in the simulated H\,{\sc ii} region. A clear ionization edge is visible. The radius at which $n(H) = n(H^+)$ matches the analytically calculated Str\"omgren radius, which is indicated by the dashed white line. The surface layer of the disk has a lower density, which moves the ionization edge to a larger distance from the star. The theoretical prediction of the Str\"omgren radius is based on the number of ionizing photons emitted and the recombination rate of electrons and protons which depends on the temperature and density of the circumstellar medium. We only consider case B recombination with the coefficient

\begin{equation}
\alpha_\mathrm{B}(T) = 2.54 \times 10^{-13} T_4^{-0.8163 - 0.0208 \ln T_4} ~~~~{\rm [cm^3\,s^{-1}]} \label{p1:eq:recomb_coef_B}
\end{equation}
by \citet{2011piim.book.....D}, with $T_4 = \frac{T}{10^4}$. The radius at which the medium becomes neutral, the Str\"omgren radius, can then be described as 
\begin{equation}
    R_{\rm S} = \left(\frac{3}{4 \pi}\right)^{1/3} \left(\frac{Q_{\rm ion}}{\alpha_{\rm B}}\right)^{1/3} n_H^{-2/3}
\end{equation}
with $n_H$ the hydrogen particle density and $Q_{\rm ion}$ the rate at which hydrogen ionizing photons are produced. In this system $Q_{\rm ion} \sim 5 \times 10^{48}$\,s$^{-1}$.   

As the ionization edge of the model and the Str\"omgren theory agree, we conclude that the hydrogen photo-ionization and recombination balance is calculated correctly in the model. 

\section{Correlations and grid behaviour}
\begin{figure*}
    \centering
    \includegraphics[width=0.7\textwidth]{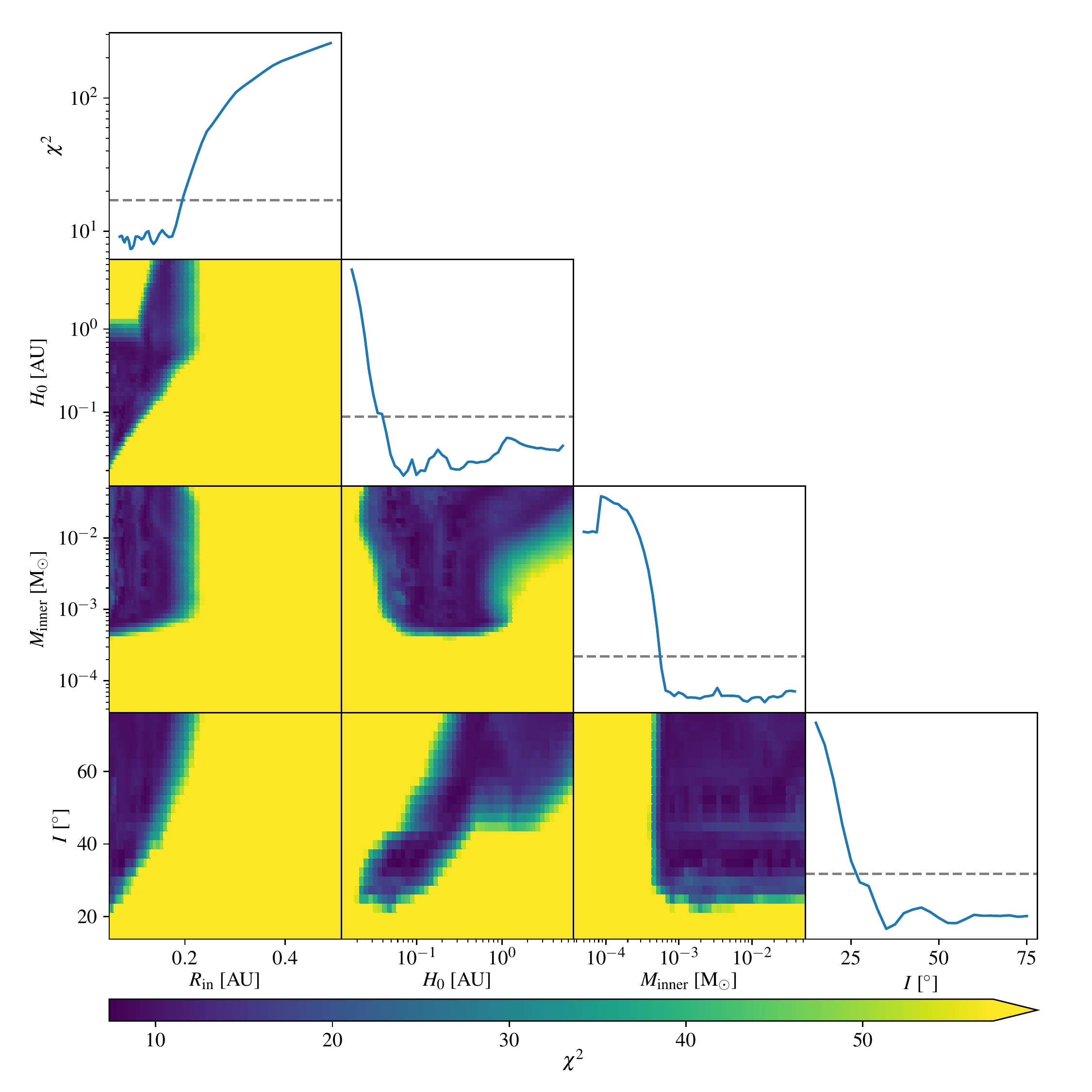}
    \caption{Corner plot indicating the $\chi^2$ value for the model fit to the Pa-12 emission line of B243 as function of each of the parameters and each parameter combination. The dashed line in each top panel indicates the 1$\sigma$ threshold.}
    \label{p1:fig:B243_correlations_Pa12}
\end{figure*}

The parameters chosen to be varied have some degeneracy, which can result in large uncertainties in their best fit value. Additionally, the coarseness of the grid can result in what we will refer to as a 'noisy' $\chi^2$ statistic. This noise is the result of the correlation between parameters and misalignment in the step size of the grid and correlation strength. This can be seen in the top panels of each column of Fig.~\ref{p1:fig:B243_correlations_Pa12} and Fig.~\ref{p1:fig:B331_correlations_Pa12}. When exploring a parameter, the optimal values of the remaining parameters can only be approximated to a limited degree. The optimal value is likely to lie between grid points. The size of this deviation varies for each point and results in a varying accuracy to which the minimum $\chi^2$ can be determined. This affects the accuracy of the optimal fit values and their uncertainties. In order to minimize this effect the coarse model grid is interpolated.

\subsection{Correlations} \label{p1:subsection:correlations}
\begin{figure*}
    \centering
    \includegraphics[width=0.8\textwidth]{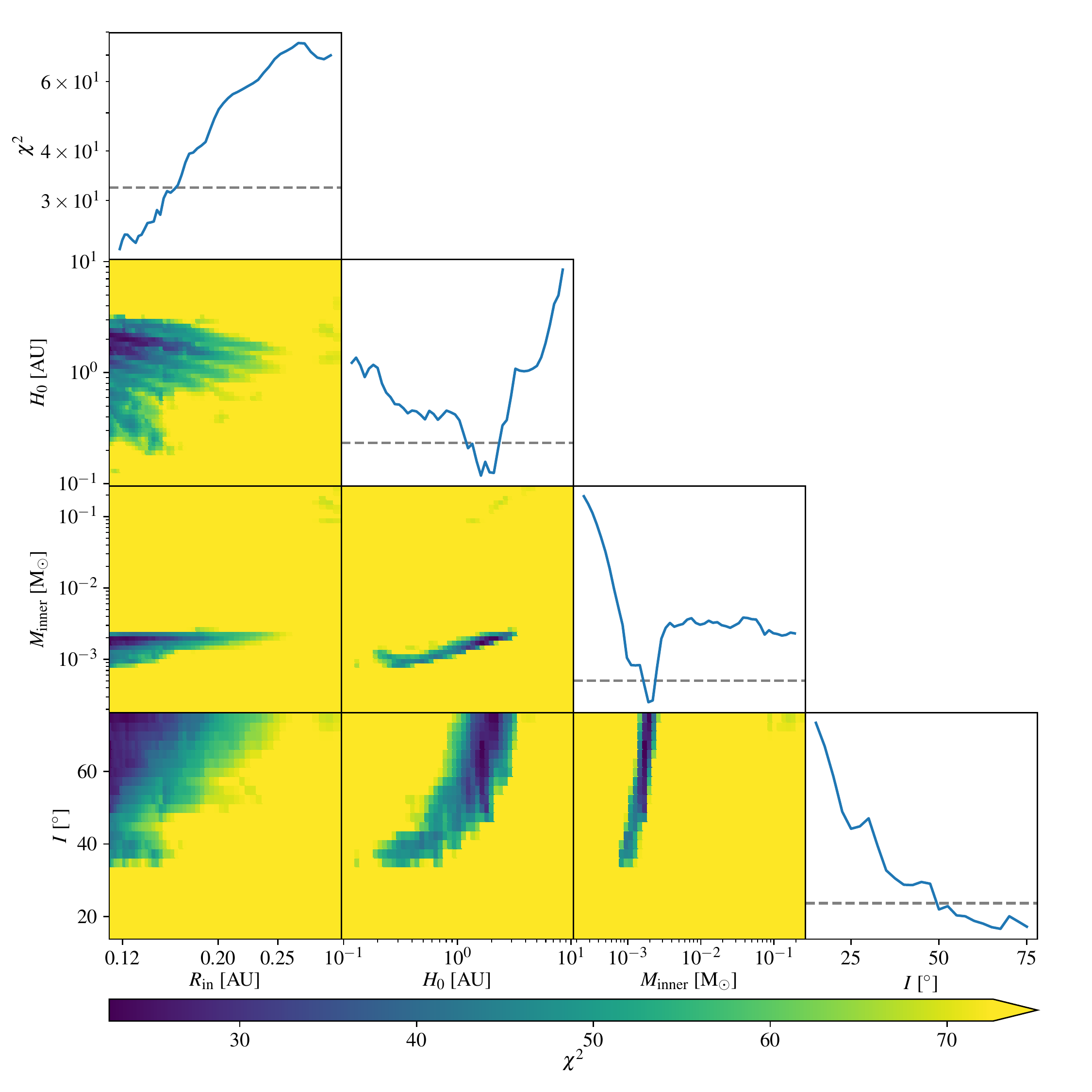}
    \caption{Corner plot indicating the $\chi^2$ value for the model fit to the Pa-12 emission line of B331 as function of each of the parameters and each parameter combination. The dashed line in each top panel indicates the 1$\sigma$ threshold.}
    \label{p1:fig:B331_correlations_Pa12}
\end{figure*}

We find correlations between the mass and scale height, inclination and scale height, and the inclination and mass as indicated in Fig.~\ref{p1:fig:B243_correlations_Pa12} and Fig.~\ref{p1:fig:B331_correlations_Pa12}. An increase in mass can result in a similar line profile by increasing the scale height and inclination of the disk. This results in degeneracy allowing a relatively large range of parameter values to fit to the data. This degeneracy combined with the coarseness of the grid results in noisy $\chi^2$ values as function of a given parameter. When the parameter value moves to a different grid point, the optimal value of another parameter might lie between two grid points and will be missed. This results in a higher $\chi^2$ value. Therefore, the accuracy of the minimum $\chi^2$ for a given parameter is limited.

\end{document}